\documentclass[a4paper,11pt]{article}
\pdfoutput=1
\usepackage{jheppub}
\usepackage[T1]{fontenc}
\usepackage{float}
\graphicspath{ {Images/} }
\title{\boldmath Doped Holographic Superconductor in an External Magnetic Field}
\author{Diego Correa,}
\author{Nicol\'as Grandi,}
\author{Alejo Hern\'andez}
\affiliation{Instituto de Fisica de La Plata - CONICET \& Departamento de Fisica - UNLP,\\C.C. 67, 1900 La Plata, Argentina}
\emailAdd{correa@fisica.unlp.edu.ar}
\emailAdd{grandi@fisica.unlp.edu.ar}
\emailAdd{alejo@fisica.unlp.edu.ar}
\abstract{We add the magnetic field axis to the holographic model for a doped superconductor proposed by Kiritsis and Li. We explore the resulting superconducting dome, in a particular region of parameters in which the equations for the superconductor can be mapped into the previously known undoped case.}
\begin{document}
\maketitle
\flushbottom
\section{Introduction}
\label{sec:Intro}
More than 30 years after its discovery,  superconductivity with high critical temperature (or ``High-$T_c$'') is still an open area of research in condensed matter physics, both from the theoretical and the experimental viewpoints (see \cite{Keimer2015FromOxides} and references therein). Being the paradigmatic cases those of cuprates and iron based superconductors, the emerging phase diagrams are very rich (see Fig.2 in \cite{Keimer2015FromOxides}). In temperature-doping phase diagrams, there is a superconducting dome in the intermediate region of the doping axis, that is known to be dominated by a $d$-wave condensate. On top of such dome, the electronic spectral function measured by ARPES experiments \cite{Damascelli2003Angle-resolvedSuperconductors} shows the presence of a Fermi surface, even if the phenomenology does not allow for a Landau description unless the doping is high \cite{Chatterjee2011ElectronicSuperconductors.}. As the doping is decreased, the Fermi surface degenerates into ``Fermi arcs'', as we move into a region known as ``pseudogap''. There, a competition of orders takes place, giving rise to inhomogeneous phases with striped and/or checkboard patterns for different order parameters. If the doping is further decreased, an anti-ferromagnetic phase appears, that in some materials is replaced by a phase with striped antiferromagnetic order parameter known as ``spin density wave'' \cite{Fradkin2015Colloquium:Superconductors}.

From a purely theoretical perspective, a class of superconducting systems can be defined in a holographic setup \cite{Gubser2008BreakingHorizon, Hartnoll2008BuildingSuperconductor}. These are known as ``holographic superconductors'' and their phenomenology shares many features with that of High-$T_c$ superconducting materials \cite{Hartnoll2009LecturesPhysics} \cite{Herzog2009LecturesSuperconductivity, Gubser2010TASISuperconductors}. Holographic realizations of the $s$-, $p$- and $d$-wave condensates have been proposed in
\cite{Sigrist1991PhenomenologicalSuperconductivity, Hartnoll2008HolographicSuperconductors, Gubser2008TheSuperconductor} and  \cite{Chen2010TowardsSuperconductors}. The fermionic spectral functions in such holographic backgrounds show the presence of a Fermi surface and, according to the parameters defining the holographic theory, regions without quasiparticles can be found that are not suitable for a Landau description \cite{Chen2010Peak-dip-humpSuperconductivity, Faulkner2010PhotoemissionSuperconductors, Gubser2010FermionSuperconductors}. The breaking of translational and/or rotational symmetry enriches the picture even further, leading to more realistic models \cite{Andrade2014ARelaxation, Ammon2019AHolography, Baggioli:2015dwa, Giataganas2012ProbingPlasma, Giataganas2018StronglyHolography}.

Most of the aforementioned theoretical research has been pursued in holographic models whose dual theory is conformal. Then, in the absence of external scales, the phase diagram can only depend on the single dimensionless quotient of temperature and chemical potential. In \cite{Kiritsis2016HolographicSuperconductivity} a holographic model was proposed, in which an additional chemical potential sets an independent scale that allows for the definition of a doping axis.  Regarding its superconducting properties, the model is very similar to the model of unbalanced superconductor first proposed in \cite{Bigazzi2012UnbalancedSpintronics} or to the model with two vector order parameters proposed in \cite{Amoretti2014CoexistenceSuperconductivity}. The introduction of an additional bulk field allows the reproduction of a phase diagram which shares many qualitative features with that of High-$T_c$ superconductors. In particular, a superconducting dome at intermediate doping and anti-ferromagnetic phase at low doping appear, as well as inhomogeneous phases in the intermediate region.

In the present work, we add a magnetic field  to the  setup of \cite{Kiritsis2016HolographicSuperconductivity}  and describe the resulting 3-dimensional phase diagrams with an additional magnetic field axis. We do so in a particular region of parameters in which the doped model can be mapped exactly into the undoped one, previously studied in \cite{Albash2008AField}.   This map is such that the model with varying doping is recast into the model without doping but with varying effective parameters, such as the values of magnetic field and scalar field charge. Then, solutions of the undoped model \cite{Albash2008AField} with varying parameters can be used to construct a phase diagram displaying both magnetic field and doping axes.

The plan of the paper is as follows. In section \ref{sec:HolographicTheory} we first present our bulk theory  with two Maxwell fields and its relevant background solution. Then, we perturb it with a charged scalar probe and map the resulting equations to the undoped model with a single Maxwell field. Later in section \ref{sec:Instability} we explore the superconducting instability both analytically and numerically. The analysis of the results is  presented in section \ref{sec:dicussion}.
\section{Holographic  model for the doped superconductor}
\label{sec:HolographicTheory}
We work with a reduced version of the model proposed in \cite{Kiritsis2016HolographicSuperconductivity}, including only the degrees of freedom involved in the superconducting transition, to lowest order.  The resulting action then reads
\begin{equation}
\label{eq:Action}
S=
\frac{1}{2\kappa_4^2} \int d^4x \; \sqrt[]{-g}
\left[
\mathcal{R} \!+\! \frac{6}{L^2}
-\frac{1}{4}F^2-\frac{1}{4}\bar{F}^2
-\lvert \partial \Psi\! -\! i q A \,\Psi \!-\! i \bar q \bar{A}\, \Psi \rvert^2
- m^2\lvert \Psi \rvert^2
\right]\,.
\end{equation}
Here  $L$ is the AdS radius,  while  $q$   and  $\bar q$ are the charges of the   massive  scalar field $\Psi$ with respect to the two Maxwell fields $A_\mu$ and $\bar A_\mu$ respectively, whose gauge curvatures are $F_{\mu\nu}$ and $\bar F_{\mu\nu}$. Notice that, as compared to the model \cite{Kiritsis2016HolographicSuperconductivity}, here we are allowing only for minimal coupling of the charged scalar to the gauge fields. This is essential to simplify our calculations in the following section by mapping the model to the undoped case,  but in principle it can be relaxed.    This model is very similar to the one proposed in \cite{Bigazzi2012UnbalancedSpintronics}, the main difference being that now the scalar field is charged with respect to both Maxwell fields.

In this model, the charged scalar $\Psi$ represents the dual of a superconducting $s$-wave order parameter. The Maxwell field $A_\mu$, which realizes a bulk gauge symmetry, corresponds to the the global $U(1)$ symmetry in the boundary theory related to particle number conservation. On the other hand, the second electromagnetic field $\bar A_\mu$ sets a scale in the holographic theory, allowing for a definition of a ``doping'' axis.  This can be loosely interpreted as related to a second particle number conservation due to the impurities of the dual theory.
\subsection{The background: a doubly charged dyonic black hole}
\label{sec:Background}
We want a background solution of the above defined dynamics, representing a normal phase in a uniform magnetic field,  in which the scalar field vanishes. A generic ansatz with transverse two dimensional rotational and translational symmetry reads
\begin{eqnarray}
&&
ds^2 = \frac{\alpha^2L^2}{z^2}\left(-fdt^2+dx^2+dy^2\right)+\frac{L^2}{z^2f}\,dz^2,
\nonumber\\
&&
 A = A_t \, dt + A_y \, dy,
\qquad \,
\bar{A} = \bar{A}_t \, dt + \bar{A}_y \, dy,
\qquad
\Psi=0\,,
\label{eq:Ansatz}
\end{eqnarray}
in which the lapse function $f$ and both gauge curvatures are assumed to depend only on the coordinate $z$.

By plugging this ansatz into the equations of motion, we get a solution in the form of a doubly charged dyonic black-hole, with a planar horizon. The lapse function and Maxwell fields are written as
\begin{eqnarray}
\label{eq:BackgroundSolution}
&&f =1-4\left(1- \frac{\pi T}{\alpha}\right)z^3+\left(3-\frac{4\pi T}{\alpha}\right)z^4,
\\
&&
A_t = \mu \left(1-z\right),
\qquad \
A_y =B\,x\, ,
\\
&&
\bar{A}_t = {\sf x}\mu \left(1-z\right),
\qquad
\bar{A}_y ={\sf y} B  \,x\,,
\end{eqnarray}
The horizon sits at $z=1$, and the AdS boundary at $z=0$. The horizon value of the gauge fields have been tuned to zero in order to have a smooth Euclidean continuation.

In the above solution, the boundary value of the $A_t$ field, given by the constant  $\mu$, represents the chemical potential of the charged particles on the boundary theory. The magnetic field acting on them is given by $B$. On the other hand, the boundary value of the field $\bar{A}_t$, given by the  constant ${\sf x}\mu$, sets a scale in the boundary theory and it can be loosely associated with the chemical potential of the impurities. This implies that the ratio ${\sf x}$   can be interpreted as  a measure of the doping. There is also an additional magnetic field ${\sf y}{B}$ acting on impurities, that represents an additional integration constant of our model. The magnitude $T$ is the temperature on the boundary theory. In order for the metric to have a smooth Euclidean continuation, it must satisfy
\begin{eqnarray}
\label{eq:Temperature}
T&=&\frac{\alpha}{4\pi}\left({3}-\frac{1}{4\alpha^4}\left(\alpha^2\mu^2(1+{\sf x}^2)+{B^2}(1+{{\sf y}^2)}\right)\right).
\end{eqnarray}

From the dual perspective, the solution presented here represents the normal state of the theory. As we   will see in the forthcoming sections, such normal state is unstable at low enough temperatures with respect to fluctuations on the scalar field. Such instability ends up in a hairy black hole solution with a non-trivial profile for the scalar field, that in the dual theory represents a superconducting phase.
\subsection{The probe: a charged scalar perturbation}
\label{sec:Probe}
If we now turn on a static perturbation of the scalar in the above background $\Psi=0+\psi$, the resulting energy momentum tensor is quadratic in the perturbation $\psi$. This implies that the induced deformation on the metric is second order, and we do not need to take it into account.  The same is true for the non-trivial electric current and the resulting deformation on the Maxwell field.

The perturbation $\psi$ satisfies the Klein Gordon equation, namely
\begin{equation}
\frac{1}{\sqrt{-g}} \, D_{\mu} \left(\,\sqrt[]{-g} \, g^{\mu\nu} D_{\nu} \right)\psi  - m^2\psi = 0 \,,
\label{eq:KleinGordon}
\end{equation}
where the gauge covariant derivative includes both Maxwell fields $D_{\mu} = \partial_{\mu} - i q A_{\mu}- i \bar{q} \bar{A}_{\mu}$.
Particularizing to the background \eqref{eq:Ansatz} we get
\begin{equation}
\alpha^2 z^2 \, \partial_{z} \left(\frac{f}{ z^2}\, \partial_{z}\psi  \right)
+
 \partial_{x}^2 \psi
+
\left(\mu^2(q+\bar{q}{\sf x})^2
\frac{(z-1)^2}{f}
- B^2(q +\bar{q}{\sf y })^2x^2
- \frac{\alpha^2m^2L^2}{z^2}\right)\psi = 0 \,,
\label{eq:KleinGordonExplicit}
\end{equation}
We solve the above equation with regular boundary conditions at the horizon, and without a source term at the AdS boundary. In consequence, if the scalar field develops a non-trivial profile, we conclude that the boundary $U(1)$ is broken spontaneously, giving rise to a superconducting phase.
\subsection{The mapping: effective parameters and the undoped case}
\label{sec:Mapping}
A key observation, that we exploit in the rest of the paper, is  the fact that  we can define effective parameters  in terms of which the problem gets mapped onto the undoped holographic superconductor in a magnetic field, first studied in \cite{Albash2008AField}. Indeed, defining effective parameters as
\begin{eqnarray}
q_{\sf eff}\,\mu_{\sf eff}&=&(q +\bar{q}\,{\sf x})\mu,
\label{eq:EffectiveParameters1}
\\
q_{\sf eff}\,B_{\sf eff}&=&(q+\bar{q}\,{\sf y})B,
\label{eq:EffectiveParameters2}
\\
T_{\sf eff}&=&T,
\label{eq:EffectiveParameters3}
\end{eqnarray}
where $T_{\sf eff}$ satisfies
\begin{eqnarray}
\label{eq:TemperatureEffective}
T_{\sf eff}&=&\frac{\alpha}{4\pi}\left({3}-\frac{1}{4\alpha^4}\left(\alpha^2\mu^2_{\sf eff}+B^2_{\sf eff}\right)\right),
\end{eqnarray}
the equation of motion of the scalar field   \eqref{eq:KleinGordonExplicit} becomes
\begin{equation}
\alpha^2 z^2 \, \partial_{z} \left(\frac{f}{ z^2}\, \partial_{z}\psi  \right)
+
 \partial_{x}^2 \psi
+
\left(
q_{\sf eff}^2\mu_{\sf eff}^2\frac{(z\!-\!1)^2}{f}
-q_{\sf eff}^2 B_{\sf eff}^2\,x^2
- \frac{m^2{
\alpha^2 L^2}}{z^2}\right)\psi = 0 \,,
\label{eq:KleinGordonMapped}
\end{equation}
where the function $f$ is now written as in \eqref{eq:BackgroundSolution} but with $T$ replaced by $T_{\sf eff}$. This is exactly the problem studied in \cite{Albash2008AField},
so in the rest of the paper we will replicate the  analysis  of that reference and map   the results  back to our system using  (\ref{eq:EffectiveParameters1})-(\ref{eq:EffectiveParameters3}).
However, the results presented in \cite{Albash2008AField}, which correspond to $q_{\sf eff} = 1$, are not enough for our purpose and we will need to extend the numerical resolution for other values of  $q_{\sf eff}$.

Borrowing from \cite{Albash2008AField} the standard procedure to deal with this equation, we separate variables with the ansatz $\psi(x,z) = X(x)Z(z)$, getting
\begin{eqnarray}
&&
Z''
+
\left(\frac{f'}{f}-\frac{2}{z}\right)Z'
+
\frac{1}{f}
\left(
q_{\sf eff}^2\mu_{\sf eff}^2\frac{(z\!-\!1)^2}{ {\alpha^2}f}
-\frac{k^2}{\alpha^2}
- \frac{m^2L^2}{z^2}\right)Z = 0 \,,
\label{eq:KleinGordonSeparatedZCrude}
\\
&&
X''
+
\left(
k^2
-
q_{\sf eff}^2 B_{\sf eff}^2\,x^2\right)X=0\,,
\label{eq:KleinGordonSeparatedXCrude}
\end{eqnarray}
where $k$ is a separation constant.

In order to solve equations \eqref{eq:KleinGordonSeparatedZCrude}-\eqref{eq:KleinGordonSeparatedXCrude} in the bulk, we impose regular boundary conditions at the horizon and at the boundary.
Close to the boundary, the variable $Z$ behaves as
\begin{equation}
Z\simeq Z_+ z^{\Delta_+}+Z_-z^{\Delta_-}\,,
\end{equation}
while close to the horizon it satisfies
\begin{equation}
Z\simeq Z_{\sf reg}+Z_{\sf div}\log(1-z)\,,
\end{equation}
Following \cite{Albash2008AField} we integrated numerically the equations starting from the horizon with a regular solution $Z_{\sf div}=0$ and shooting with the parameter $\mu_{\sf eff}$ so as to get $Z_-=0$ at the boundary. Whenever a non-trivial profile for the scalar field exists, the $U(1)$ symmetry of the boundary theory is spontaneously broken and the system becomes superconducting.

In the following section, we analyze the planes $(T,{\sf x})$, $(T,B)$ and $(B,{\sf x})$ separately, and then construct the full phase diagram by using the above defined mapping from the $(T_{\sf eff},B_{\sf eff})$ plane into the $(T,B,{\sf x})$ space.
\section{The superconducting instability}
\label{sec:Instability}
\subsection{Finite temperature and doping, and vanishing magnetic field}
\label{sec:FiniteTemperatureInstability}
In the case of vanishing magnetic field, equations \eqref{eq:KleinGordonSeparatedZCrude}-\eqref{eq:KleinGordonSeparatedXCrude} decouple and the equation for $X$ can be solved trivially with a constant profile. The numerical solution of the equation for $Z$    leads to a superconducting region in the plane $(T,{\sf x})$. As described in  \cite{Kiritsis2016HolographicSuperconductivity}, depending on the values of parameters $m$, $q$ and $\bar q$, different sorts of phase diagrams corresponding to different sorts of  phenomenology are obtained.

\begin{figure}[ht!]
\centering
\includegraphics[scale=.7]{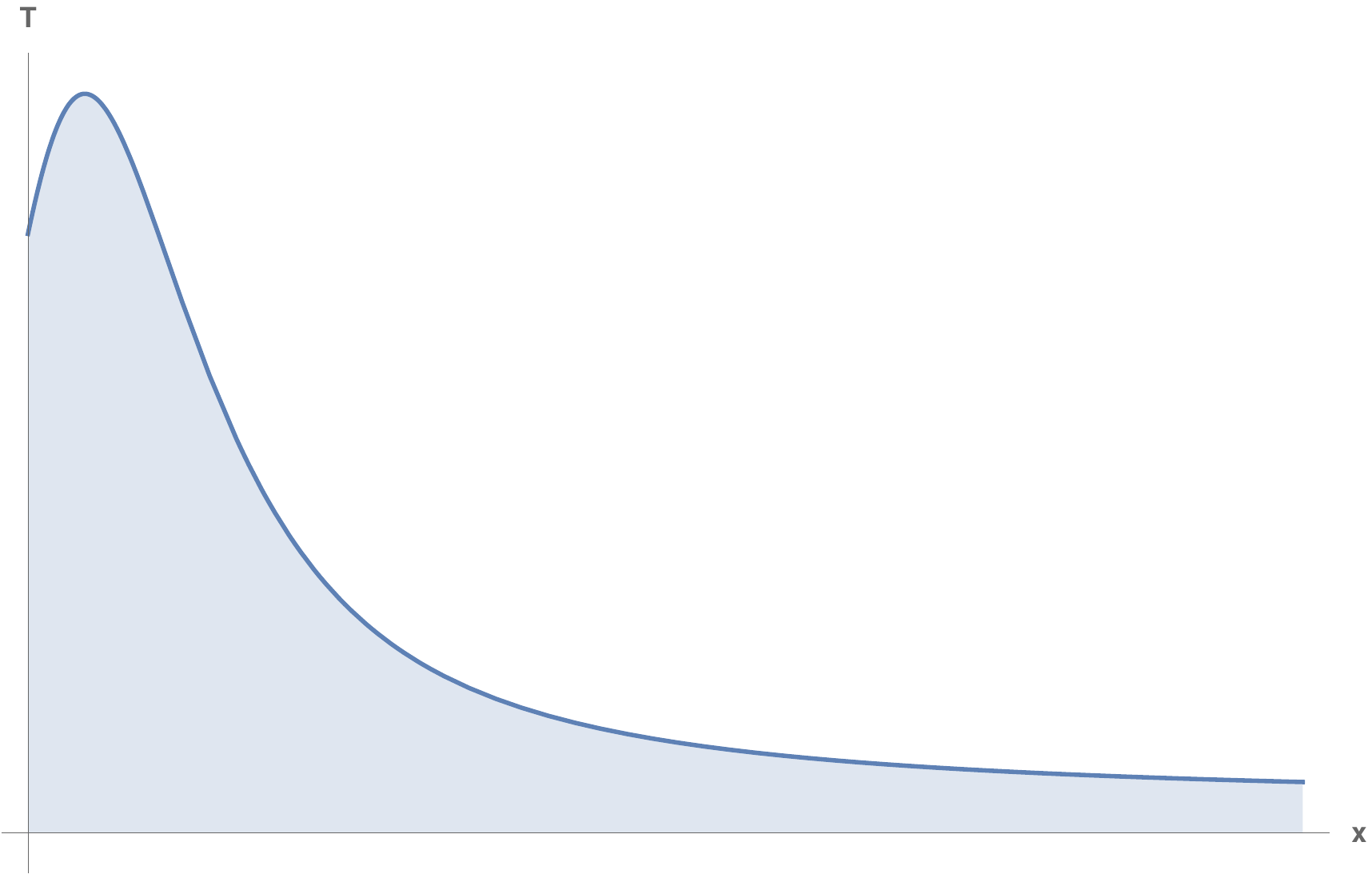}
\caption{Generic phase diagram in the $(T,{\sf x})$ plane.}
\label{fig:generic-kiritsis}
\end{figure}

In the particular example  depicted in Fig. \ref{fig:generic-kiritsis} the resulting phase diagram   presents a ``dome'' which extends for all positive values of ${\sf x}$. However, depending on the case, the superconducting phase may exist only for finite range of ${\sf x}$. A generic  behaviour for the different cases we shall discuss, is that the critical temperature exhibits a maximum value at a finite value of the doping ${\sf x}$.

In order to prove that analytically, we notice that when the magnetic field $B$ vanishes, the effective magnetic field $B_{\sf eff}$ vanishes as well. Then, the regularity condition at the AdS boundary $Z_-=0$ imposes a relation between the two remaining free parameters of the effective model $Z_-[q_{\sf eff}, \mu _{\sf eff}]=0$, which  can be solved as $q_{\sf eff}=p\left(\mu_{\sf eff}/\sqrt{12}\right)$, where $p(\cdot)$ is a function that has to be determined numerically, and the factor $\sqrt {12}$ is included for later convenience. From \eqref{eq:TemperatureEffective}, the critical temperature reads
\begin{equation}
T_c=\frac{3\alpha}{4\pi}\left(1-\left(p^{-1}(q_{\sf eff})\right)^2\right)
\label{eq:Teff.pmenosuno}
\end{equation}
where $p^{-1}(\cdot)$ stands for the inverse of $p(\cdot)$. When $B=0$, the mapping \eqref{eq:EffectiveParameters1}-\eqref{eq:EffectiveParameters3}  can be solved explicitly for
\begin{equation}
q_{\sf eff} =\frac{q+\bar q{\sf x}}{\sqrt{1+{\sf x}^2}},
\label{eq:q.eff}
\end{equation}
and equation \eqref{eq:Teff.pmenosuno} can be used to study the properties of the critical temperature as a function of the doping.

In particular, the maximum value of the critical temperature sits at a value of the doping such that
\begin{equation}
\partial_{\sf x}T_c = -\frac{3\alpha}{2\pi}\frac{p^{-1}(q_{\sf eff})}{p'(\mu_{\sf eff}/\sqrt{12})}\,\partial_{\sf x}q_{\sf eff}= 0.
\label{eq:stationary.points}
\end{equation}
Solving $\partial_{\sf x}q_{\sf eff}=0$ we obtain a value of ${\sf x}$ at which the critical temperature is stationary
\begin{equation}
{\sf x}_{\sf max}=\frac{\bar q}q   .
\end{equation}
Additional stationary points might exist if the prefactor in \eqref{eq:stationary.points} vanished at some particular values of $q_{\sf eff}$, whose position in the doping axis could then be obtained from \eqref{eq:q.eff}.

\begin{figure}[H]
\centering
\includegraphics[scale=.6]{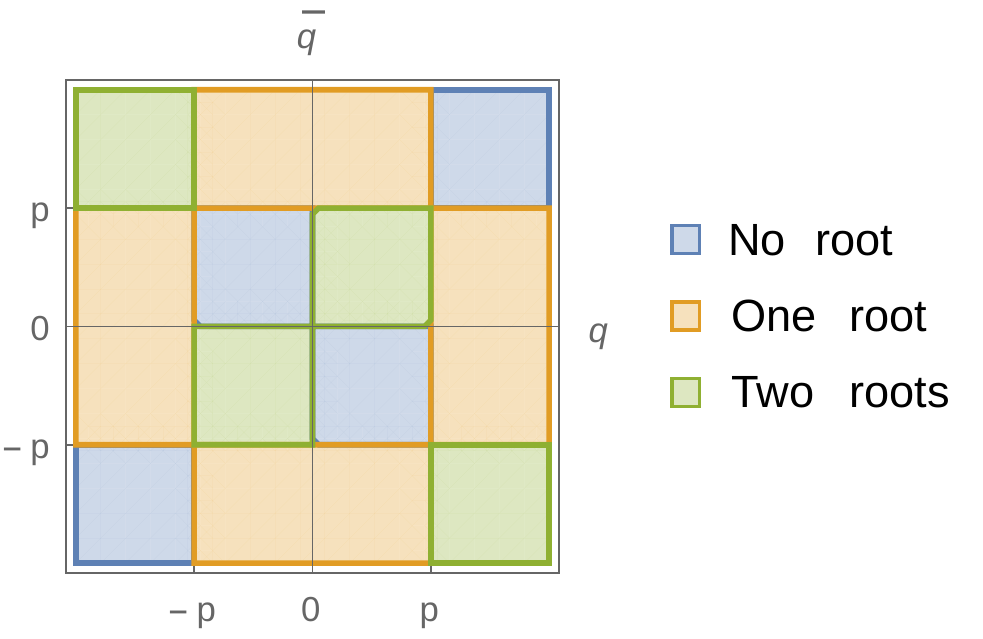}
\caption{The $(q,{\bar q})$ plane, showing the regions in which there are no, one or two roots for the critical temperature as a function of ${\sf x}$.
The critical value $p$ corresponds to $p_\pm$ (or to, $p_0$ see bellow)
}
\label{fig:tateti}
\end{figure}

On the other hand, the critical temperature \eqref{eq:Teff.pmenosuno} vanishes at the values of ${\sf x}$ that satisfy $q_{\sf eff}=p(\pm 1)\equiv p_\pm$, the sign being that of $\mu_{\sf eff}$. This can be rewritten as
\begin{equation}
\left(p_\pm^2-\bar q^2\right){\sf x}^2-2\,q\bar q\,{\sf x}+(p_\pm^2-q^2)=0.
\label{eq:tateti}
\end{equation}
In this equation, the signs of the independent and quadratic coefficients depend on $q$ and $\bar q$ respectively. Then, using Descartes' sign rule, we can draw a diagram in the $(q,\bar q)$ plane with the number of positive roots, Fig.\ref{fig:tateti}. There we see that, according to the values of $q$ and $\bar q$, we can have superconductivity along the whole doping axis (no positive root), up to a maximum doping (one positive root) or between a minimum and a maximum doping (two positive roots).

\subsection{Finite temperature and magnetic field, and vanishing doping}
\label{sec:ZeroDopingInstability}
Now we  turn to the problem of solving equations \eqref{eq:KleinGordonSeparatedZCrude}-\eqref{eq:KleinGordonSeparatedXCrude} for a nontrivial profile of the fields $Z$ and $X$ at vanishing doping.  As before,  we identify the existence of such solution with the onset of a superconducting phase.

The procedure employed to find the superconducting region in a $(T,B)$ diagram at zero doping (or equivalently in a $(T_{\sf eff},B_{\sf eff})$ diagram) is completely analogous to that of reference \cite{Albash2008AField}, as follows. We first solve equation \eqref{eq:KleinGordonSeparatedXCrude}  for $X$ in terms of confluent hypergeometric functions with $k$ initially unconstrained. Then, we assume a finite profile for $X$, which constrains $k$ to be proportional to an odd integer, {\em i.e.}, $k = \sqrt{q_{\sf eff} B_{\sf eff}}\,(2n+1)$. This has the effect of turning the confluent hypergeometric functions into Hermite polynomials of order $n$, namely, $H_n$. Finally, as $\psi$ phase must remain constant in order to satisfy the equation of motion, we are forced to choose $n$ to be zero, because $H_0$ is the only Hermite polynomial that is an even function.
Having obtained the allowed value $k=\sqrt{q_{\sf eff}B_{\sf eff}}$, we solve numerically equation \eqref{eq:KleinGordonSeparatedZCrude} by means of a shooting method.
We do this for several discrete values of $q_{\sf eff}$ and $m$.

\begin{figure}[ht!]
\centering
\includegraphics[scale=.7]{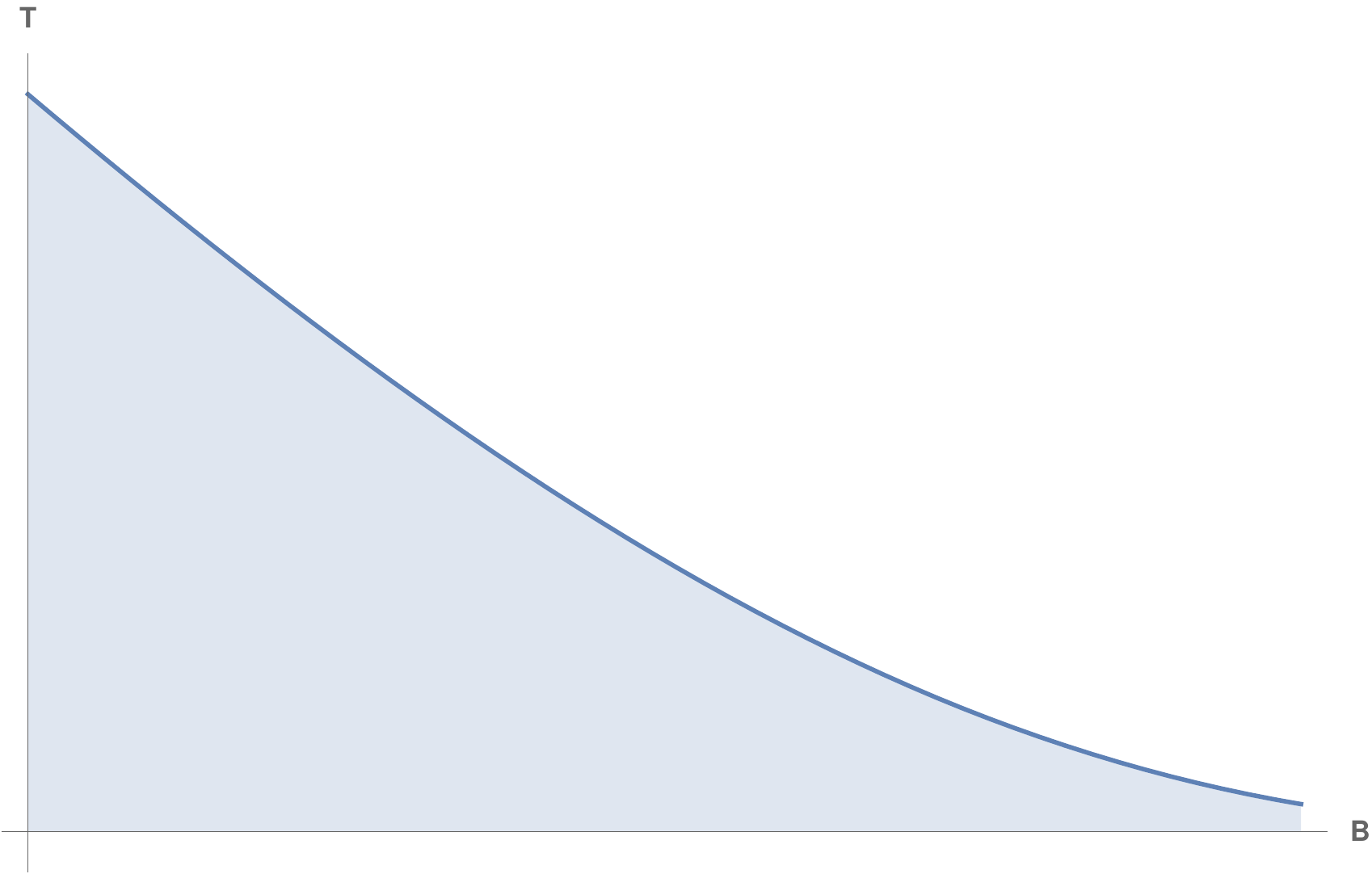}
\caption{Generic phase diagram in the $(T,B)$ plane.}
\label{fig:generic-johnson}
\end{figure}

Depending on the values of $q_{\sf eff}$ and $m$, a superconducting region may exist or not. For the cases in which it does exist, the phase diagram is generically as the one depicted in Fig.\ref{fig:generic-johnson}.
\subsection{Vanishing temperature, and finite magnetic field and doping}
\label{sec:ZeroTemperatureInstability}
When the Beckenstein-Hawking temperature of the black hole \eqref{eq:Temperature} vanishes, the metric has a near horizon expansion with the form
\begin{equation}
ds^2= \alpha^2L^2_{(2)}\left(-\frac{1}{\zeta^2} dt^2+dx^2+dy^2\right)
+
\frac{L^2_{(2)}}{\zeta^2}d\zeta^2
\label{eq:MetricNearHorizon}
\end{equation}
Where we defined $\sqrt{6}(z-1)=1/\zeta$.
This corresponds to an $AdS_2\times \mathbb{R}^2$ geometry, with $AdS_2$ radius given by $L_{(2)}=L/\sqrt{6}$.

The scalar equations \eqref{eq:KleinGordonSeparatedZCrude}-\eqref{eq:KleinGordonSeparatedXCrude} in this limit reduce to
\begin{eqnarray}
&&
\zeta^2Z''
-
m_{(2)}^2
L_{(2)}^2Z = 0 \,,
\label{eq:KleinGordonSeparatedNearHorizonZ}
\\
&&
X''
+
q_{\sf eff}B_{\sf eff}\left(
1
-
q_{\sf eff}B_{\sf eff}x^2\right)X=0\,.
\label{eq:KleinGordonSeparatedNearHorizonX}
\end{eqnarray}
where now the prime denotes a derivative with respect to $\zeta$. Equation \eqref{eq:KleinGordonSeparatedNearHorizonZ} corresponds to that of a scalar field fluctuation  on the $AdS_2$ background defined by \eqref{eq:MetricNearHorizon} with a mass
\begin{equation}
m^2_{(2)}L^2_{(2)}
=
\frac{1}{6}\left(
m^2L^2
+ \frac{ q_{\sf eff}B_{\sf eff}}{\alpha^2}
-
\frac{1}{6\alpha^2}q_{\sf eff}^2\mu_{\sf eff}^2
\right).
\label{eq:MassAdS2}
\end{equation}
In order for $AdS_2$ background to be stable  under the scalar field fluctuation, the mass $m^2_{(2)}L^2_{(2)}$ must satisfy the Breitenlohner-Freedman bound in two dimensions
\begin{equation}
m^2_{(2)}L^2_{(2)}>-\frac14.
\label{eq:MassAdS2BF}
\end{equation}
The violation of this bound would indicate an instability of the near-horizon black hole geometry,  {\em i.e.} the development of a superconducting phase. In terms of the physical parameters of the boundary theory, the above condition reads
\begin{equation}
m^2L^2+\frac32
+
(q+\tilde{q}{\sf y})\frac{B}{\alpha^2}
+
\frac{(q+\tilde{q}{\sf x})^2}{1+{\sf x}^2}
\left(
(1+{\sf y}^2)\frac{B^2}{6\alpha^4} - 2 \right)
>0.
\label{BreitelohnerFreedmanAdS2}
\end{equation}
We expect the condensation of our system whenever the left hand side becomes negative.

\begin{figure}[h]
\centering
\includegraphics[scale=.7]{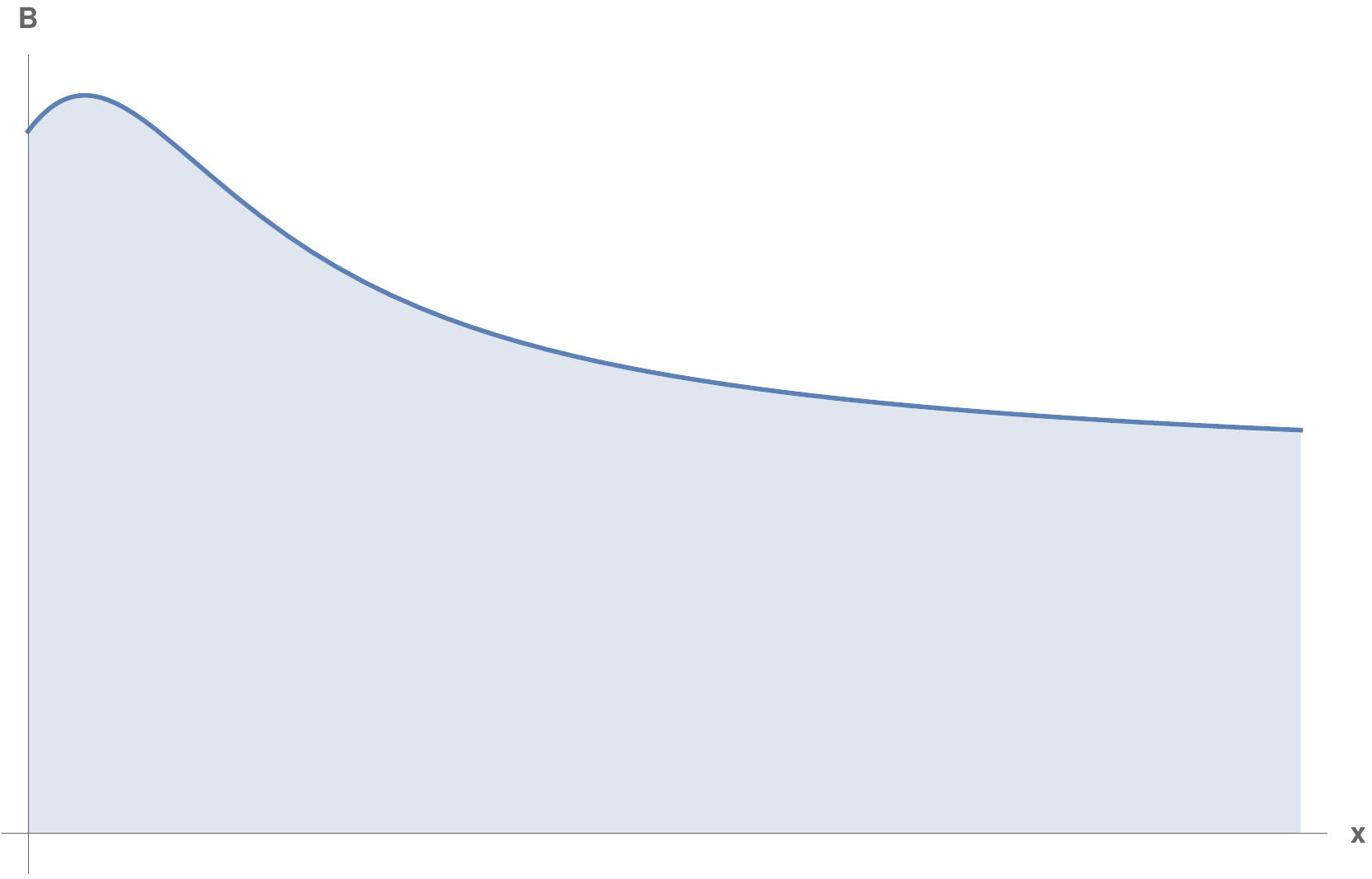}
\caption{Generic phase diagram in the $(B,{\sf x})$ plane.}
\label{fig:generic-breitelohner}
\end{figure}

A generic phase diagram for  a suitable choice of parameters is shown in Fig.\ref{fig:generic-breitelohner}. We see that the superconducting phase disappears into the normal phase whenever the magnetic field $B$ is large enough. The maximum value of the magnetic field, that allows for a superconducting phase, depends on the doping.

 In order to understand this analytically, we saturate the bound putting $B=B_c$ and rewrite it as
\begin{equation}
\left(p_0^2-\bar q^2\right){\sf x}^2-2\,q\bar q\,{\sf x}+(p_0^2-q^2)=0,
\label{eq:bound}
\end{equation}
with
\begin{equation}
p_0=\frac{m^2L^2+\frac32+
(q+\bar q{\sf y})\frac {B_c}{\alpha^2}}{2-(1+{\sf y}^2)\frac{B_c^2}{6\alpha^2}}.
\label{eq:p0}
\end{equation}
At $B_c=0$, {\em i.e.} on the ${\sf x}$ axis, we get $p_0= (m^2L^2+3/2 )/2$, and the above equation has the same structure as eq.\eqref{eq:tateti}, implying a diagram similar to that of Fig.\ref{fig:tateti} in the $(q,\bar q)$ plane. Then again, according to the values of $q$ and $\bar q$, we have superconductivity along the whole doping axis, up to a maximum doping, or between a minimum and a maximum doping.

Regarding the maximum critical magnetic field, we can compute $\partial_{\sf x}B_c$ using the implicit function theorem as follows
\begin{equation}
\partial_{\sf x}B_c=\frac{{\sf x}(p_0^2-\bar q^2)-\,q\bar q}{p_0\left(1+x^2\right)\partial_Bp_0}=0.
\end{equation}
Combined with \eqref{eq:bound}, this equation gives the result ${\sf x}_{\sf max}=\bar q/q$. At such value of the doping, the critical magnetic field is maximum. Remarkably, this value coincides with the value of the doping that maximizes the critical temperature. The resulting maximum critical magnetic field $B_c$ at ${{\sf x}={\sf x}_{\sf max}}$ can then be solved from equations \eqref{eq:bound} and \eqref{eq:p0} to obtain its explicit dependence on $m^2$, $q$, $\bar q$ and ${\sf y}$.

\subsection{The complete phase diagram}
\begin{figure}[ht!]
\centering
\includegraphics[scale=.62]{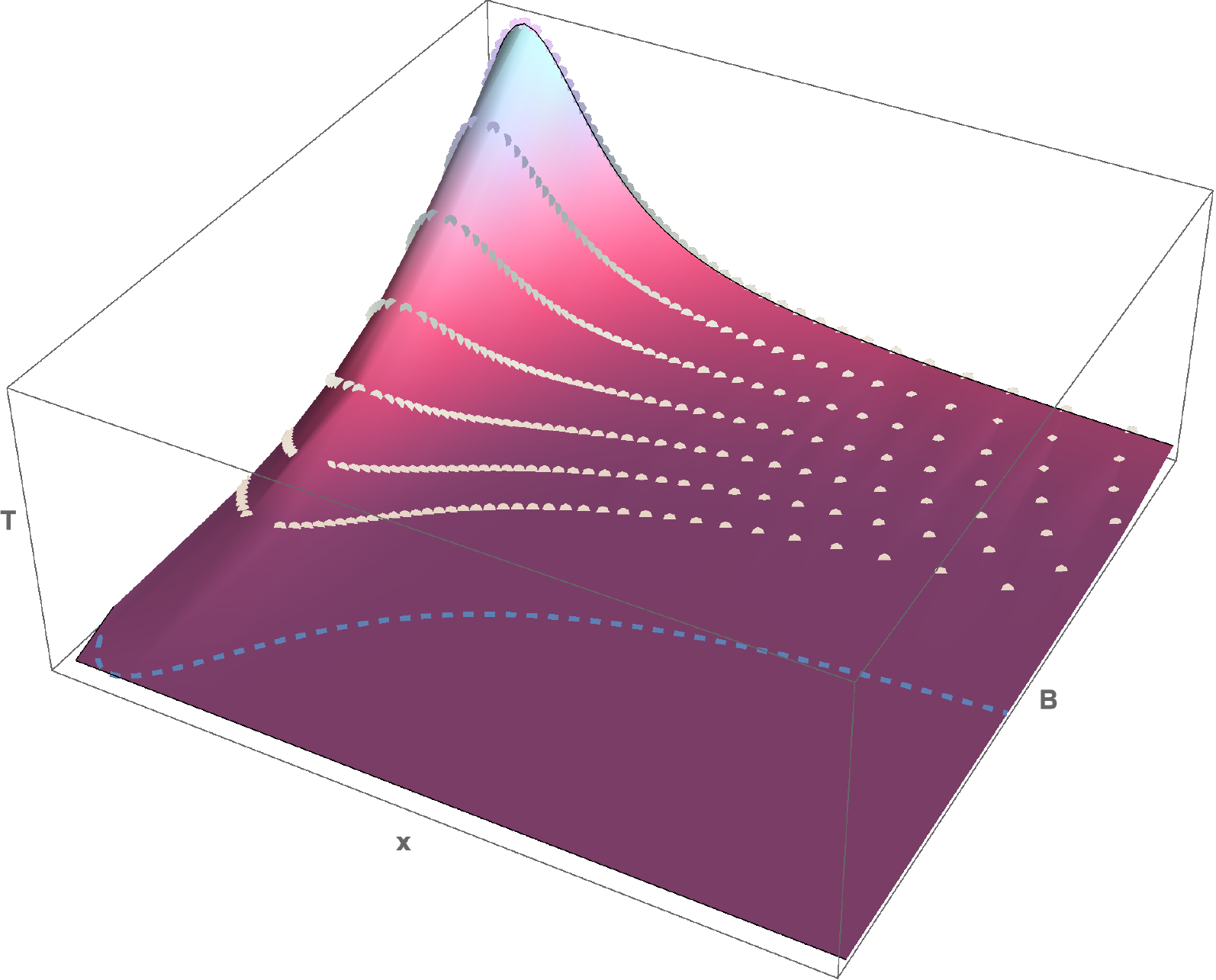}
\caption{Generic phase diagram in the $(T,B,{\sf x})$ space. The dotted curves are associated with a value of $B_{\sf eff}$.}
\label{fig:todo-generic}
\end{figure}

Zero doping solutions written in terms of the effective parameters, can be mapped to curves  in a 3-dimensional diagram with axes $(T,B,{\sf x})$ and fixed ${\sf y}$, through (\ref{eq:EffectiveParameters1})-(\ref{eq:EffectiveParameters3}). Mapping many solutions with different effective parameters we obtain a family of curves that eventually specifies a surface in the  3-dimensional phase diagram. The surface shown in Fig.\ref{fig:todo-generic} is a generic example of it, where the superconducting dome extends  non-trivially into the magnetic field axes, sitting on top of the region of the $(B,{\sf x})$ plane in which the Breitenlohner-Freedman bound is violated.

\begin{figure}[h]
\centering
\includegraphics[width=4.8cm]{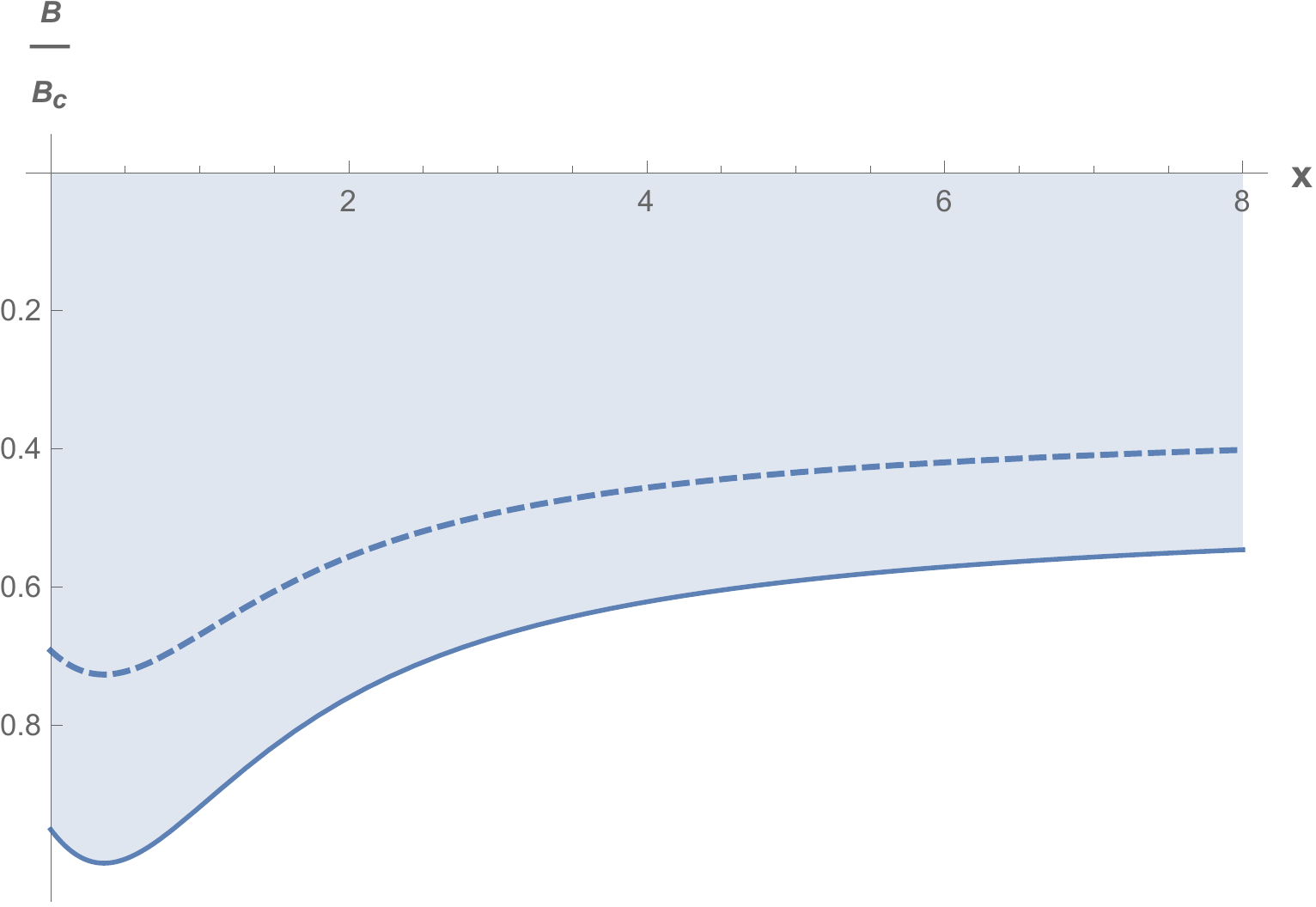}
\includegraphics[width=4.8cm]{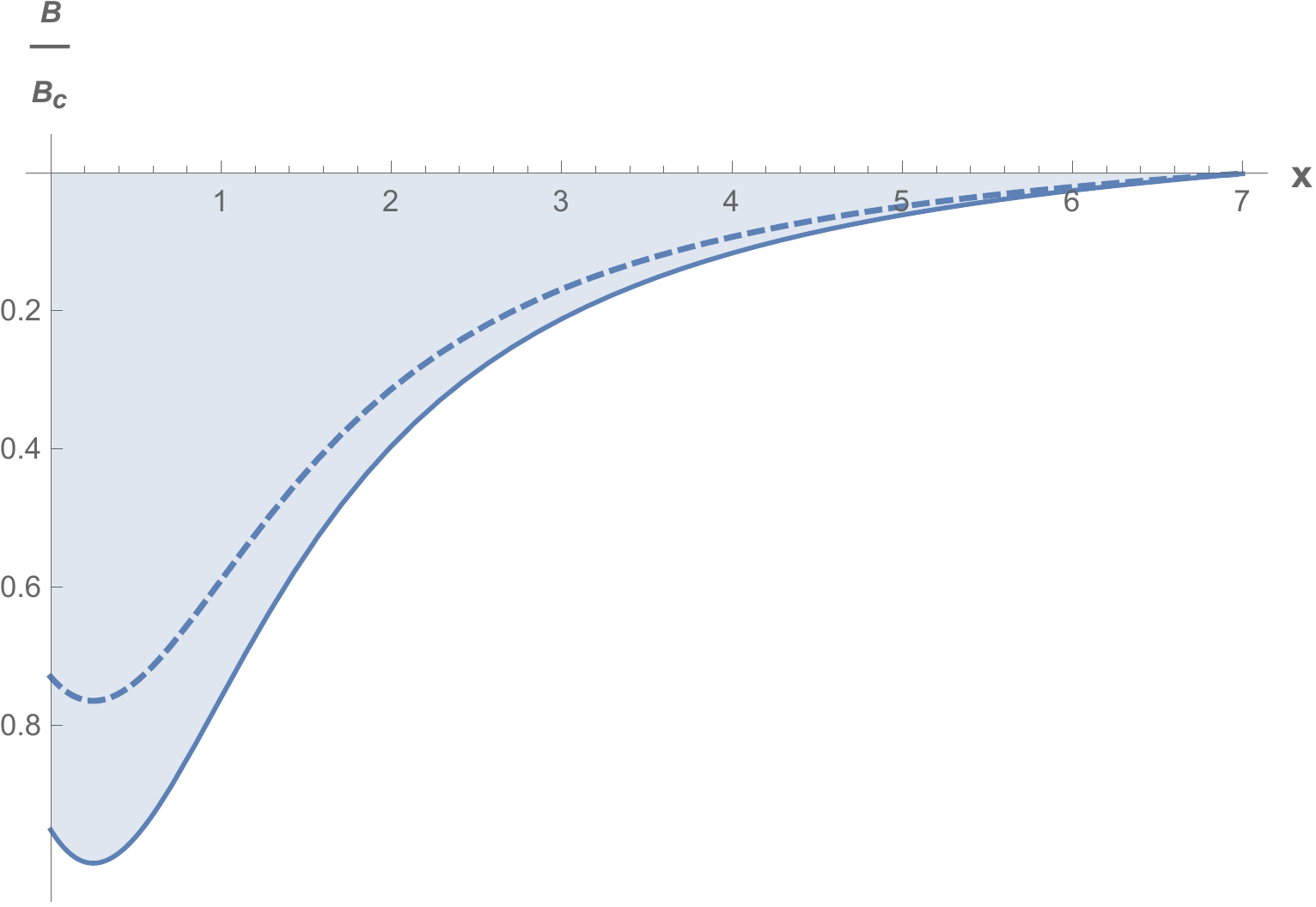}
\includegraphics[width=4.8cm]{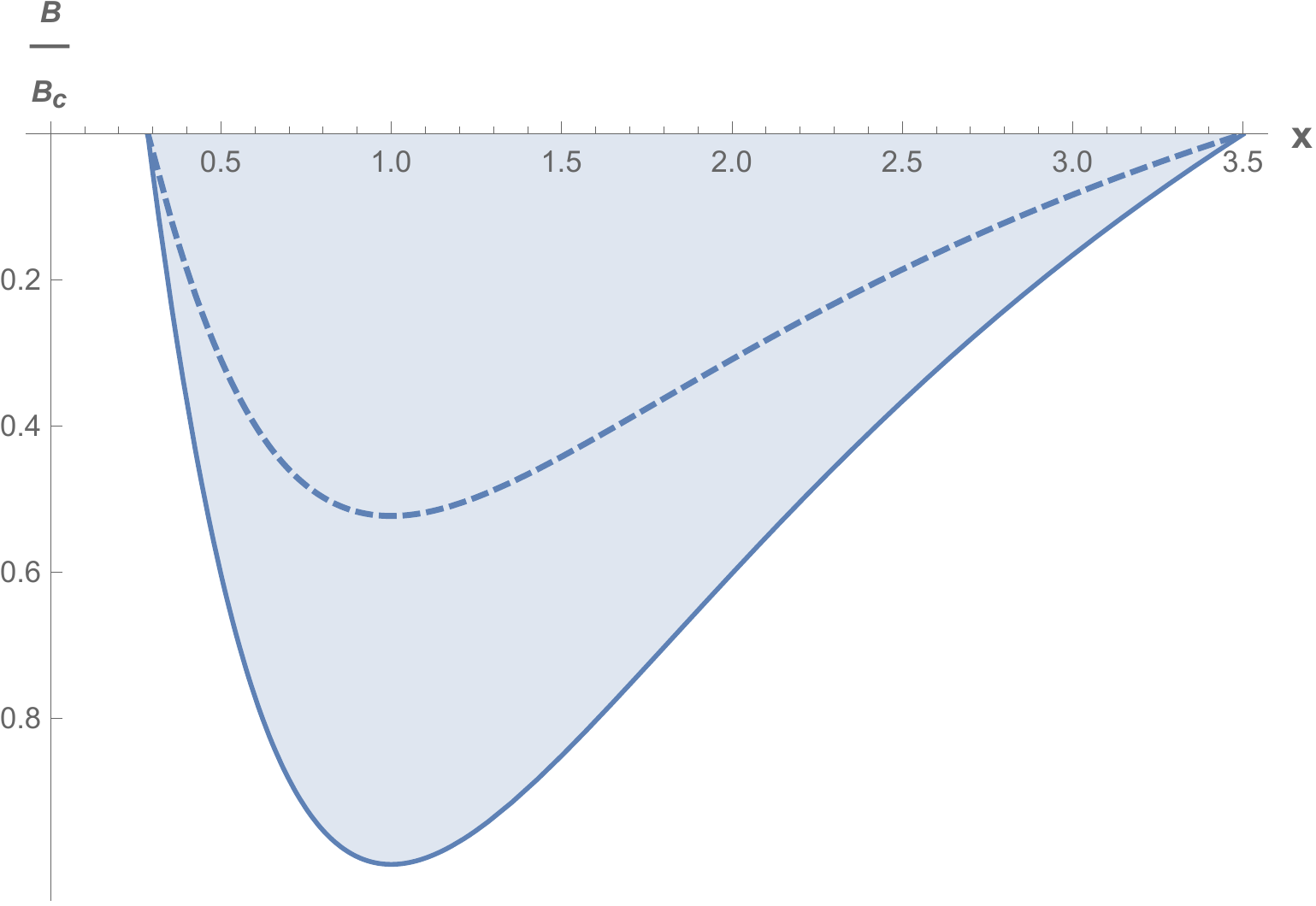}
\caption{The $T=0$ plane for the cases
 {\it i}) $q=0.7$ , $\bar q=0.25$ and $m^2=-2 $ (the superconducting region covers the whole ${\sf x}$ axis), {\it ii}) $q=1$, $\bar q=0.25$ and $m^2= -1.2 $ (there is a maximum doping up to which the superconducting phase could exist) and
 {\it iii}) $q=0.6$, $\bar q=0.6$ and $m^2= -0.4$ (the superconducting phase extends in a bounded region of the positive doping axis). 
 Dotted and solid lines correspond to the ${\sf y}=1$ and ${\sf y}=0$ cases respectively.
Plots are normalized with the maximum critical magnetic field for the ${\sf y}=0$ case.}
\label{fig:Bxplanes}
\end{figure}

The precise details of the phase diagrams depend on the specific values of $q$, $\bar q$ and $m^2$. The main distinctive feature among different cases is the range of  the parameter ${\sf x}$ in which  the superconducting region extends. We shall explore the three particular possibilities: {\it i}) the dome extends for all positive values of ${\sf x}$, {\it ii}) the dome extends from ${\sf x}=0$ up to a finite positive value of ${\sf x}$, {\it iii}) the dome extends from a minimum positive value of $ {\sf x}$ up to a maximum one. This can be easily seen in the $T=0$ plane, using the Breitenlohner-Freedman bound
\eqref{BreitelohnerFreedmanAdS2}, as in Fig. \ref{fig:Bxplanes}. Notice that the value of $ {\sf y}$ does not affect substantially the form of the superconducting region.

For non-vanishing magnetic field, we can extend our analysis of the maximum critical temperature as follows. The regularity condition at the AdS boundary $Z_-[q_{\sf eff}, \mu _{\sf eff}, B_{\sf eff}]=0$ imposes a relation on the free parameters of the effective model. Taking the total differential of this relation, we have
\begin{equation}
\partial_{q_{\sf eff}}Z_-\,dq_{\sf eff}
+\partial_{\mu_{\sf eff}}Z_-\,d\mu_{\sf eff}
+
\partial_{B_{\sf eff}}Z_-\,dB_{\sf eff}
=
0
\label{eq:diff}
\end{equation}
Now taking the total differential of
\eqref{eq:EffectiveParameters1}-\eqref{eq:EffectiveParameters3} at constant magnetic field, we have
\begin{eqnarray}
\mu_{\sf eff}\,dq_{\sf eff}+q_{\sf eff}\,d\mu_{\sf eff}
&=&\bar{q}\mu\,d{\sf x}+(q +\bar{q}\,{\sf x})d\mu,
\label{eq:EffectiveParameters1diff}
\\
B_{\sf eff}\,dq_{\sf eff}+q_{\sf eff}\,dB_{\sf eff}&=&0,
\label{eq:EffectiveParameters2diff}
\\
\alpha^2\mu_{\sf eff}\,d\mu_{\sf eff}
+
B_{\sf eff}\,dB_{\sf eff}&=&0
\end{eqnarray}
were we used the fact that at the maximum critical temperature, the temperature is stationary $dT=0$, that can be rewritten as
\begin{eqnarray}
\mu\, d\mu(1+{\sf x}^2)+\mu^2 \,{\sf x}\, d{\sf x}&=&
0.
\label{eq:tempdiff}
\end{eqnarray}
Combining \eqref{eq:diff}-\eqref{eq:tempdiff} we get the relations
\begin{eqnarray}
\alpha^2\mu_{\sf eff}\,
\left(
q_{\sf eff} \partial_{q_{\sf eff}}Z_-
-
\left(
\partial_{B_{\sf eff}} Z_-
+
\partial_{\mu_{\sf eff}}Z_-
B_{\sf eff}\right)B_{\sf eff}\right)dq_{\sf eff}
&=&0
\label{1stLine}
\\
\frac{1}{\alpha^2\mu_{\sf eff}}\left(\alpha^2\mu^2_{\sf eff}
-B_{\sf eff}\right)\,dq_{\sf eff}
&=&
\frac{\bar{q}-q\,{\sf x}}{(1+{\sf x}^2)} \,\mu\,{\sf x}\,
d{\sf x},
\label{2ndline}
\end{eqnarray}
The equation \eqref{1stLine} is satisfied whenever $dq_{\sf eff}=0$. This implies the vanishing of \eqref{2ndline}, which in turn requires that
${\sf x}={\sf x}_{\sf max}=\bar q/q$. Remarkably, such value is the same as the one we obtained at zero magnetic field, and also coincides with the position of the maximum critical magnetic field at zero temperature. This result is independent of ${\sf y}$, implying that the qualitative features of the phase diagram are robust with respect to changes in such parameter.
Again, additional stationary points might exist if the prefactor in \eqref{1stLine} vanished.
\section{Results and discussion}
\label{sec:dicussion}
\begin{figure}[H]
\centering
\includegraphics[width=6.0cm]{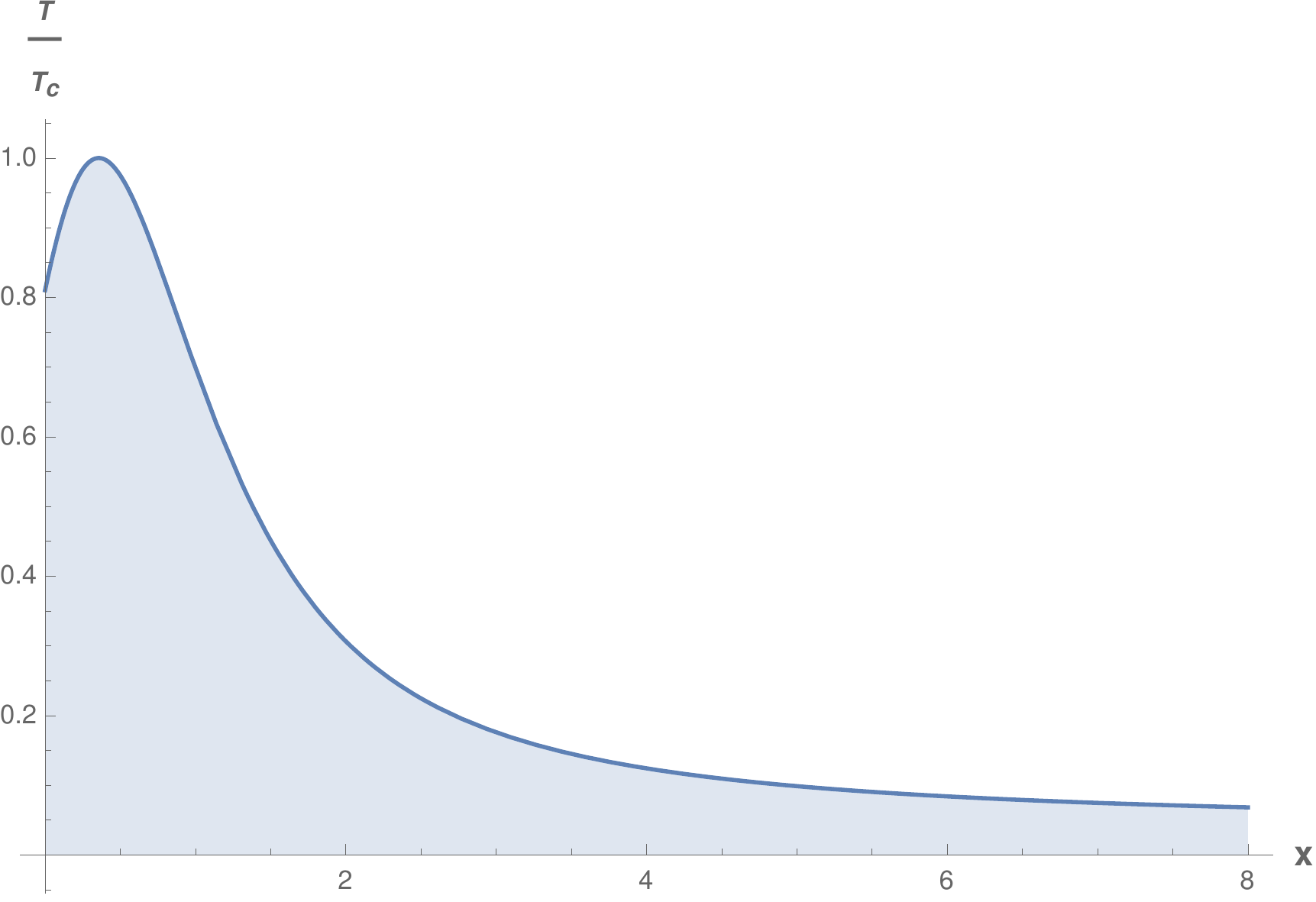}\
\includegraphics[width=6.0cm]{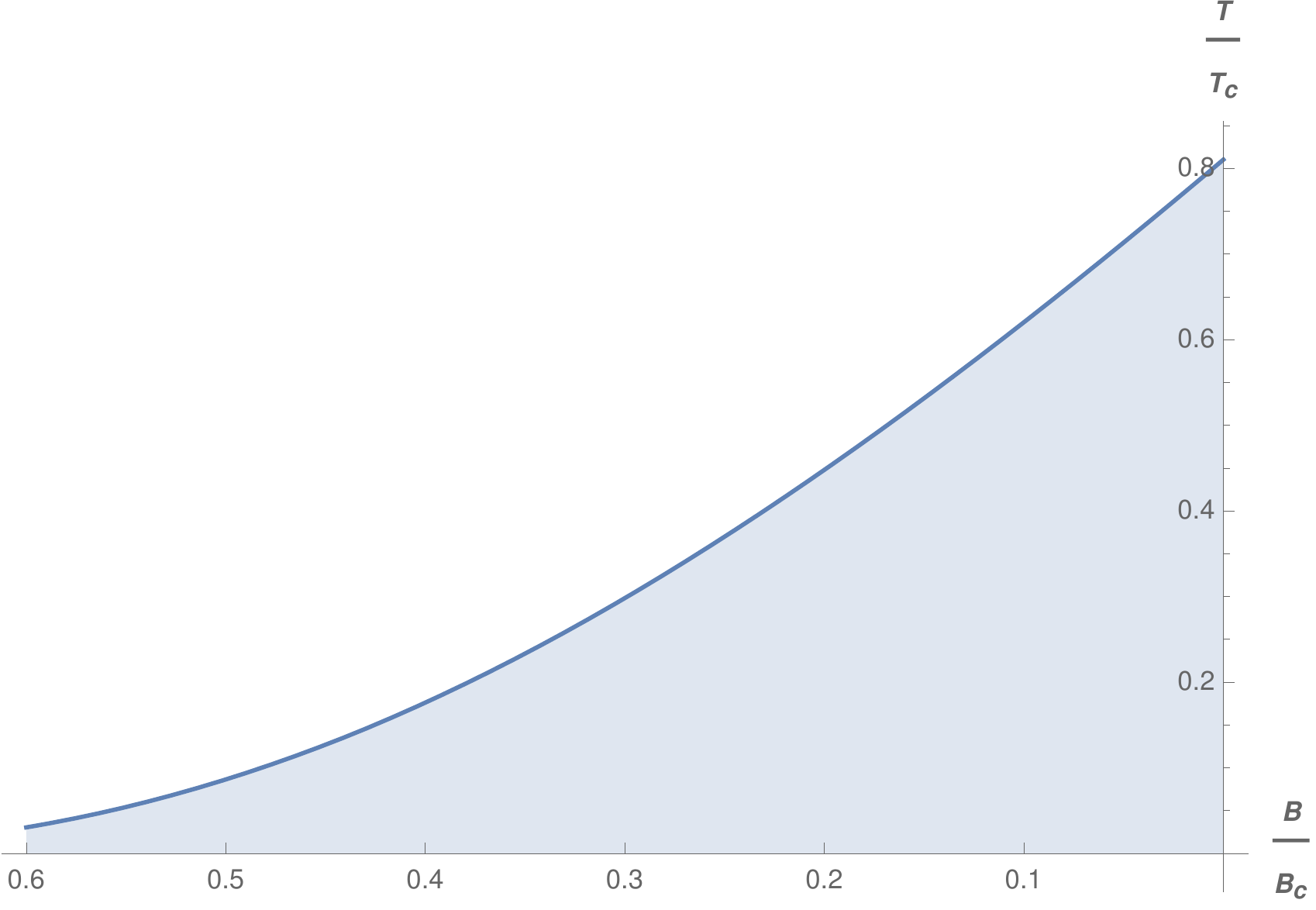}
\\
\includegraphics[width=6.0cm]{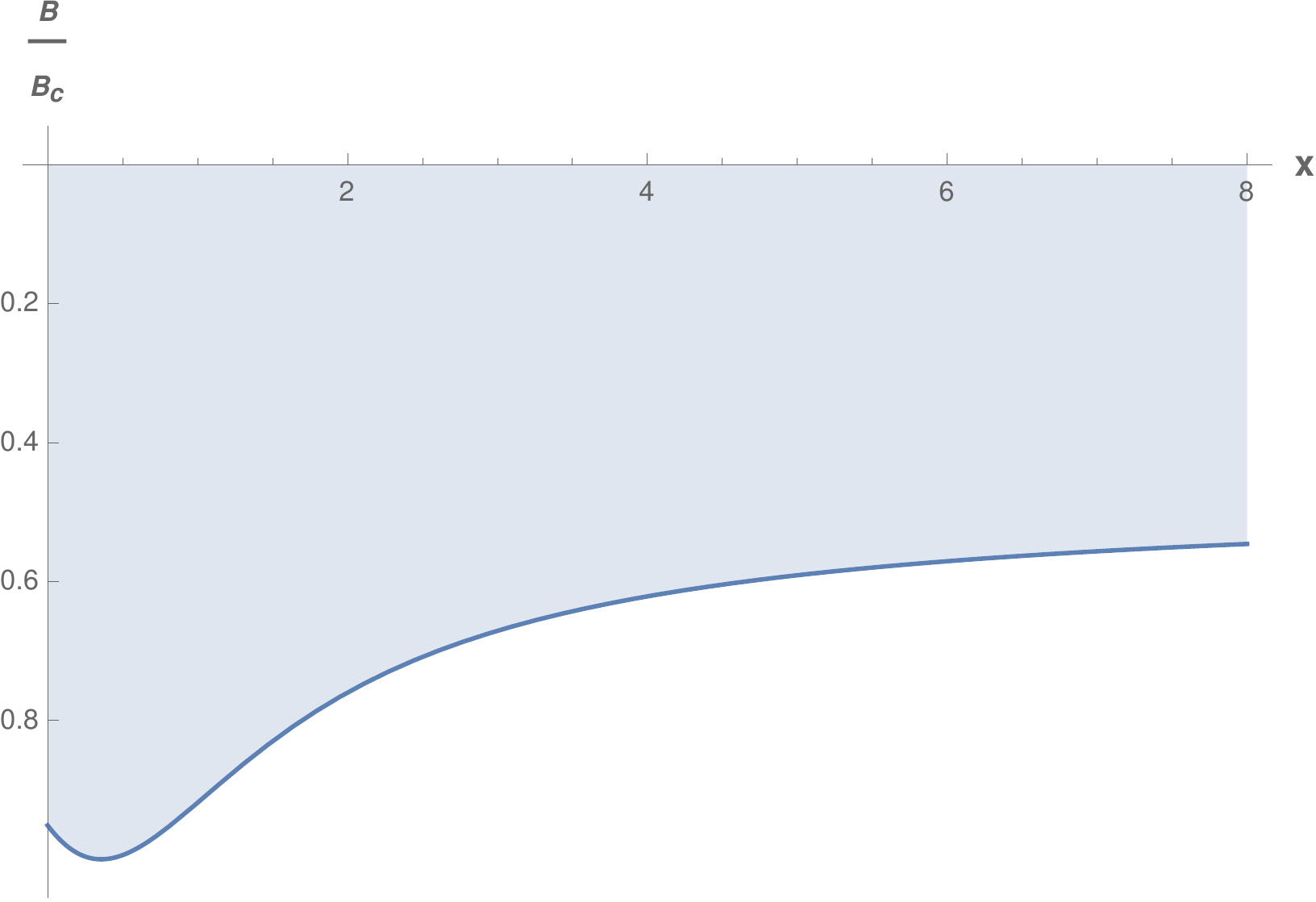}\
\includegraphics[width=6.0cm]{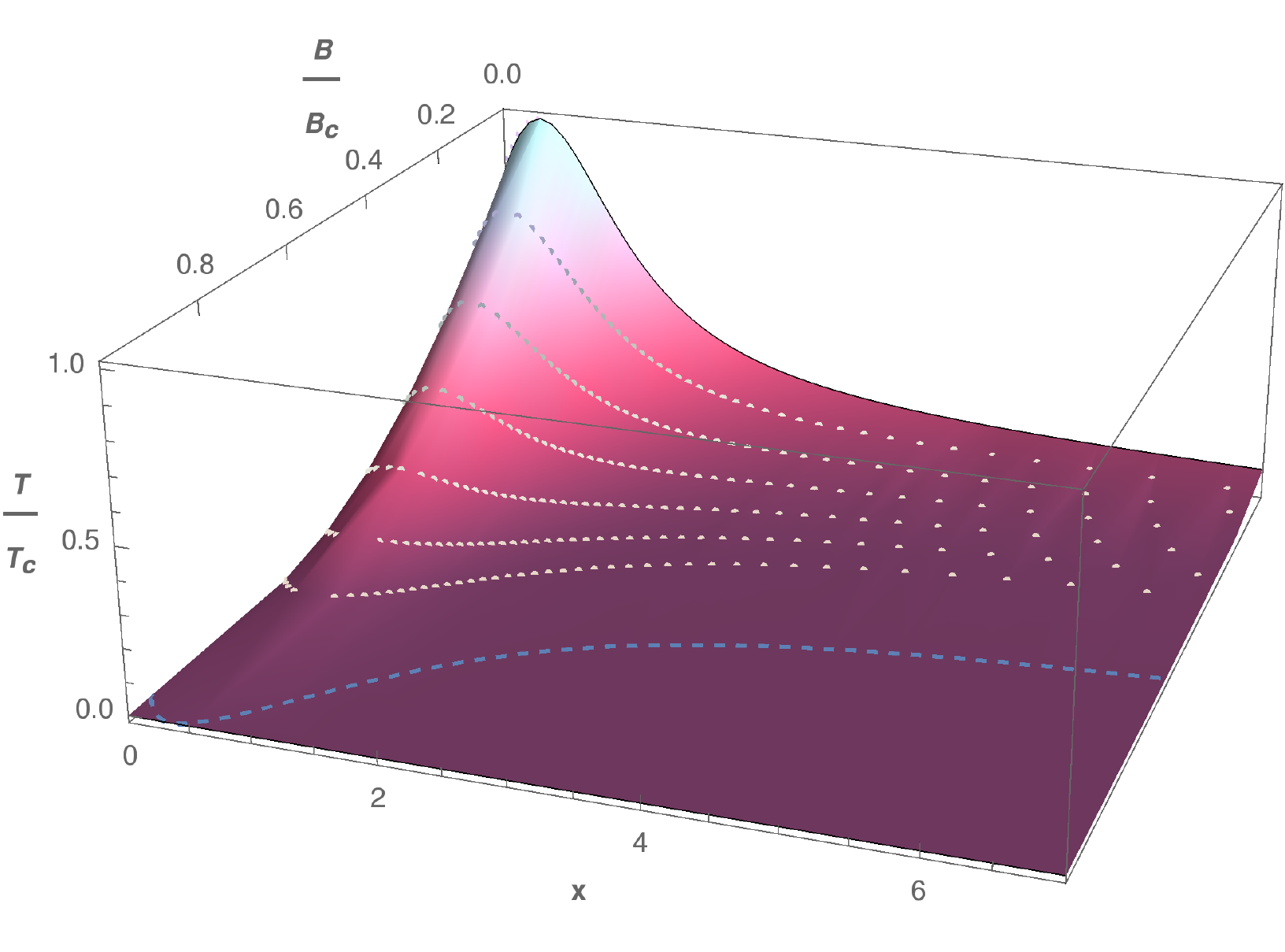}
\caption{Top: $(T,{\sf x})$ plane (left), and $(T,B)$ plane (right). Bottom: $(B, {\sf x})$ plane (left) and the complete $(T,B,{\sf x})$ diagram (right). We have $m^2=-2$, $q=0.7, \bar{q}=0.25$ 
and ${\sf y}=0$. We are normalizing the plots with $T_c$, the maximal critical temperature for $B=0$, and $B_c$, the maximal magnetic field for $T=0$. In this case $B_c \simeq 1.425 \times 10^3 T_c^2 $.
}
\label{fig:PhaseDiagram_m2=-2}
\end{figure}

\begin{figure}[H]
\centering
\includegraphics[width=6.0cm]{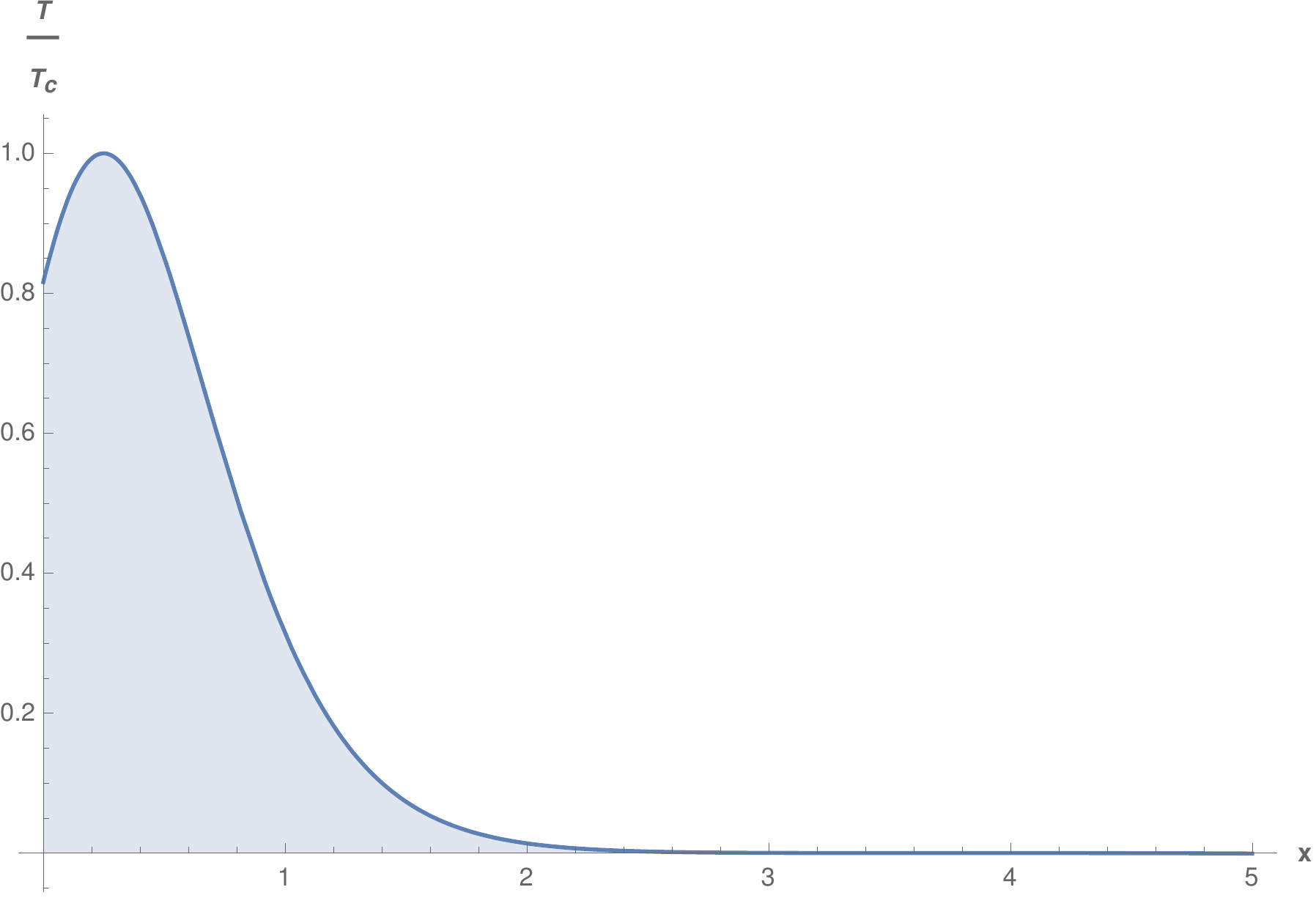}\
\includegraphics[width=6.0cm]{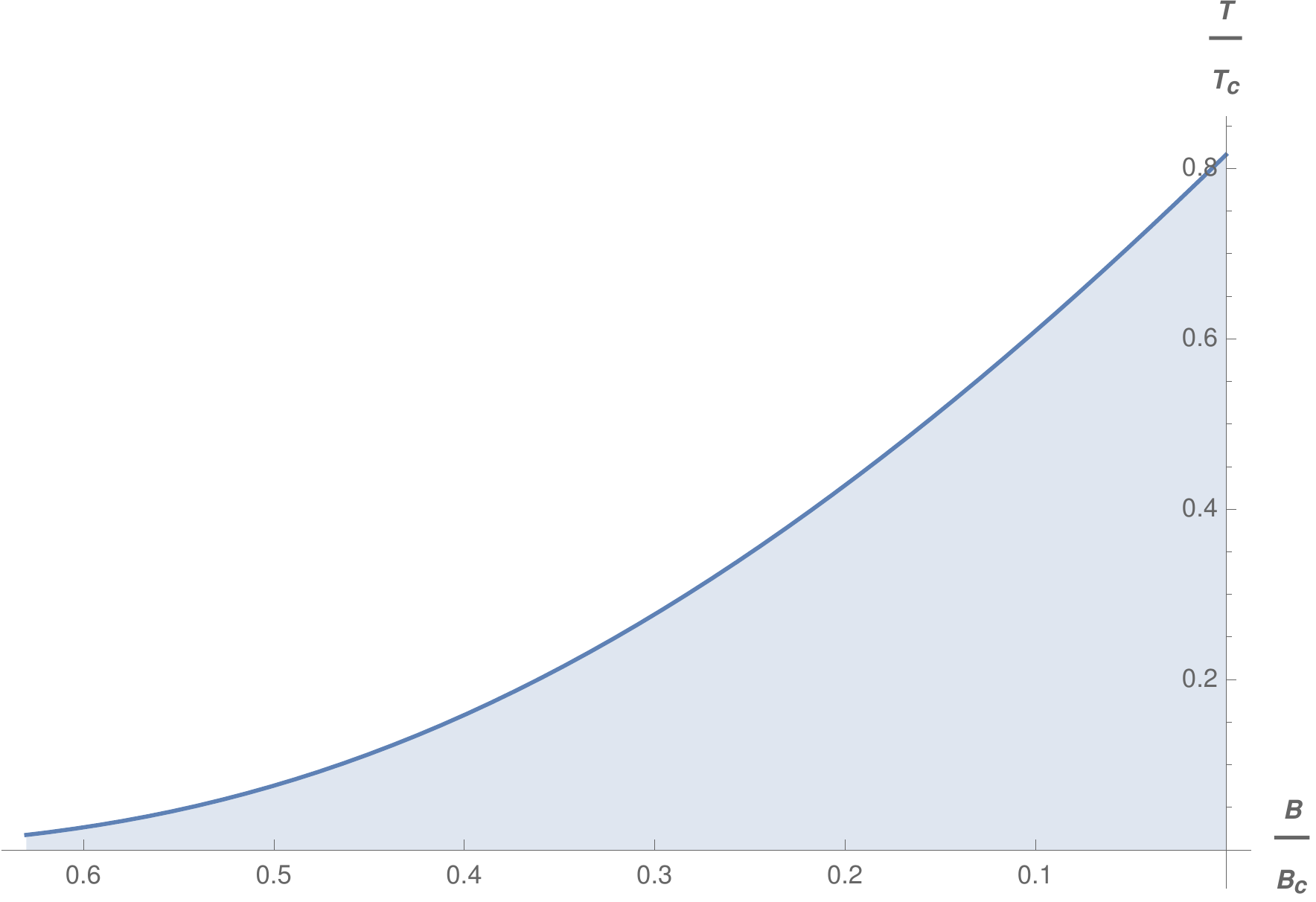}
\\
\includegraphics[width=6.0cm]{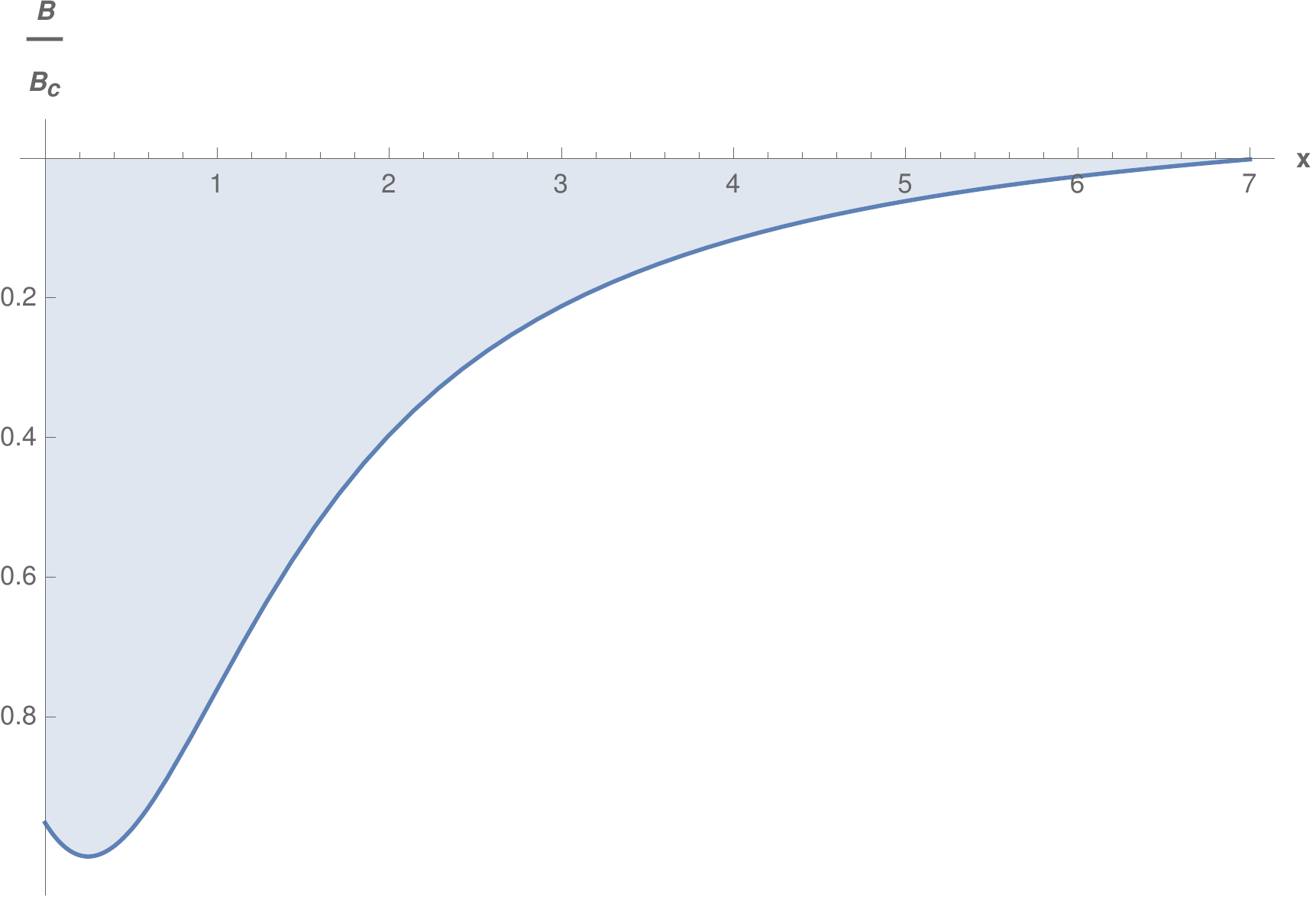}\
\includegraphics[width=6.0cm]{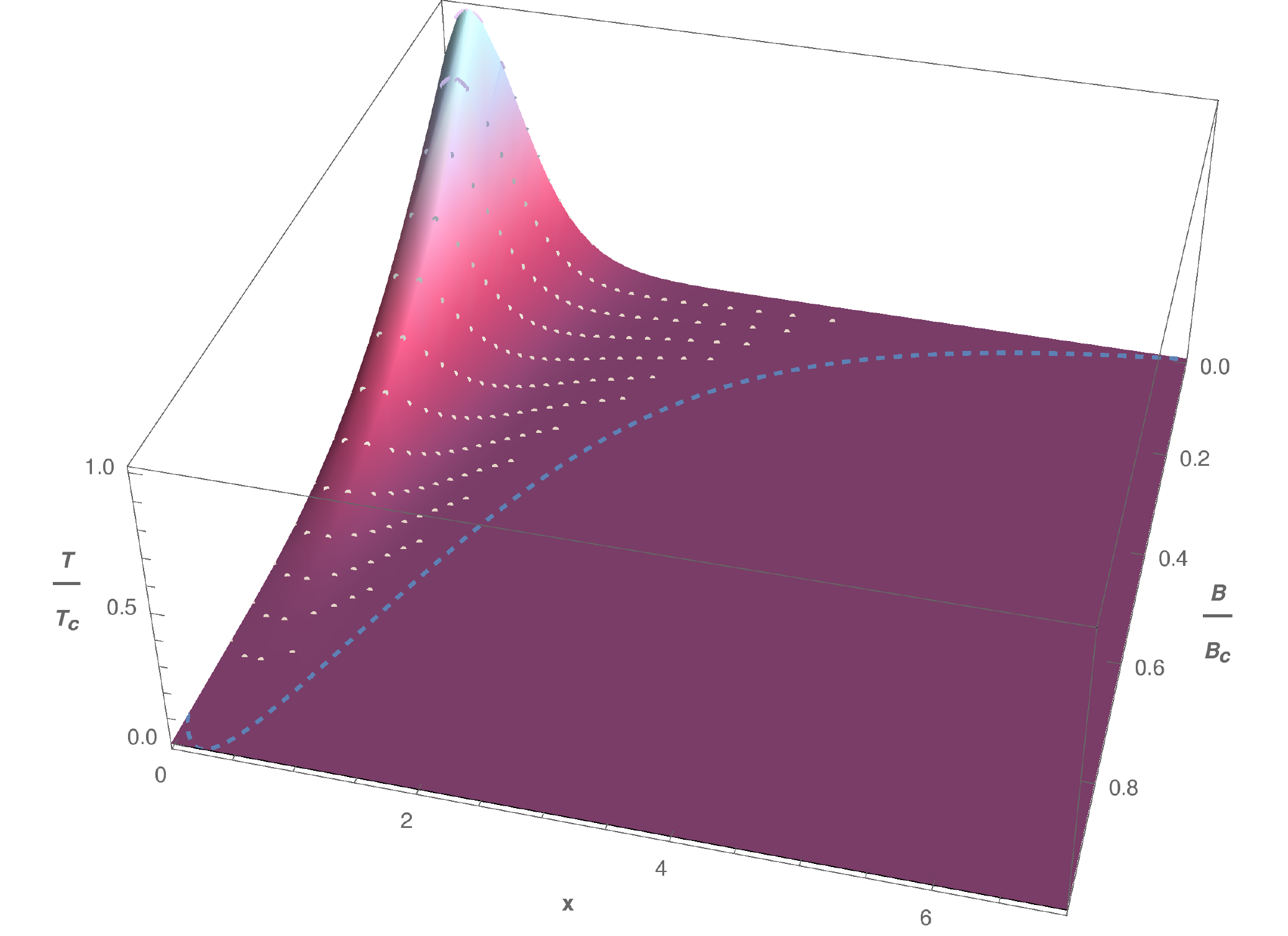}
\caption{Top: $(T,{\sf x})$ plane (left), and $(T,B)$ plane (right). Bottom: $(B,{\sf x})$ plane (left) and the complete $(T,B, {\sf x})$ diagram (right). We have $m^2=-1.2$, $q=1, \bar{q}=0.25$
and ${\sf y}=0$. We are normalizing the plots with $T_c$, the maximal critical temperature for $B=0$, and $B_c$, the maximal magnetic field for $T=0$. In this case $B_c \simeq  6.828 \times 10^3 T_c^2 $.
}
\label{fig:PhaseDiagram_m2=-12}
\end{figure}

\begin{figure}[H]
\centering
\includegraphics[width=6cm]{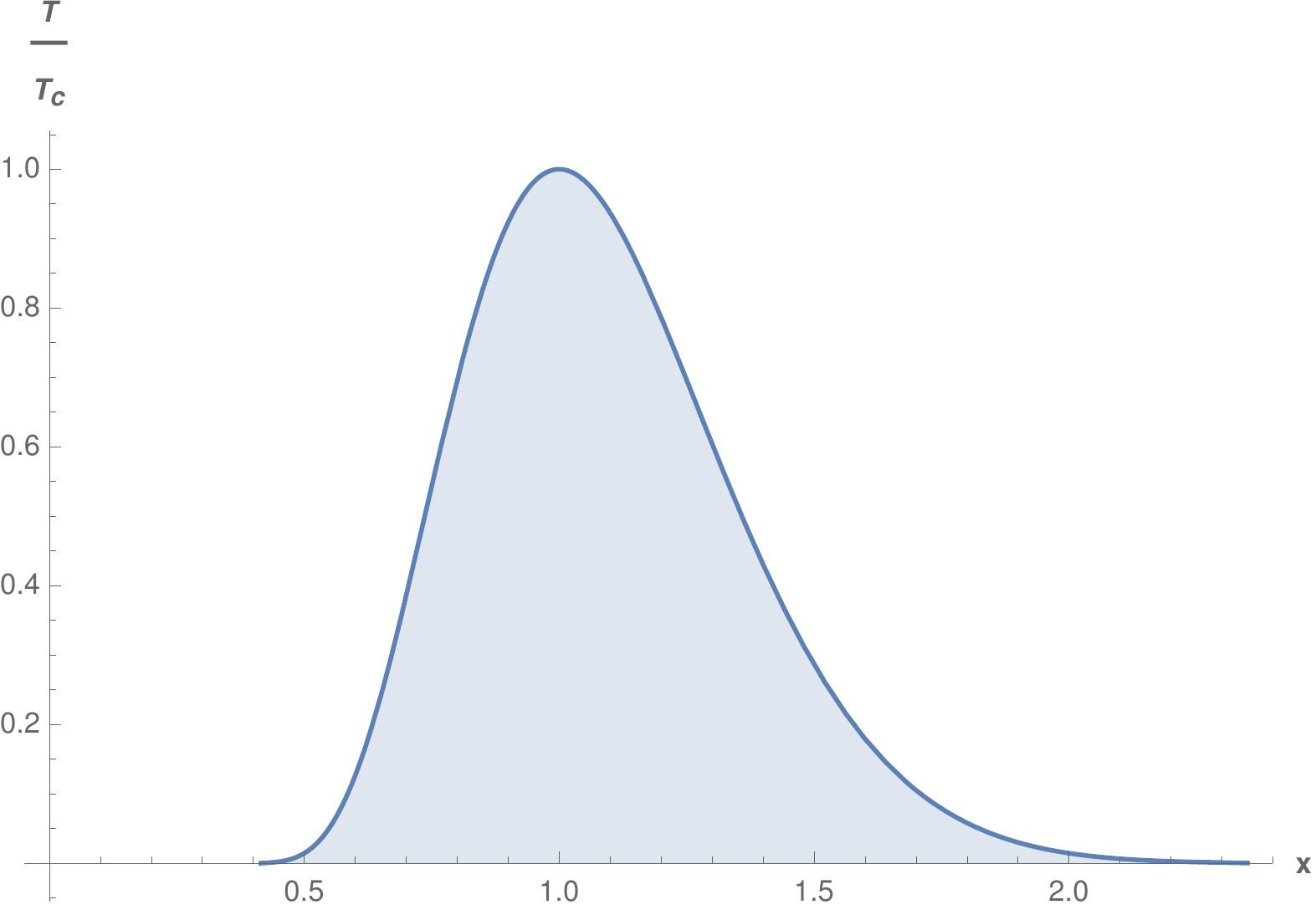}
~~~~~~~~~~~~~~~~~~~~~~~~~~~~~~~~~~~~~~~~~~~~~~~~~~~~~~~~~~~~~~~
\\
\includegraphics[width=6cm]{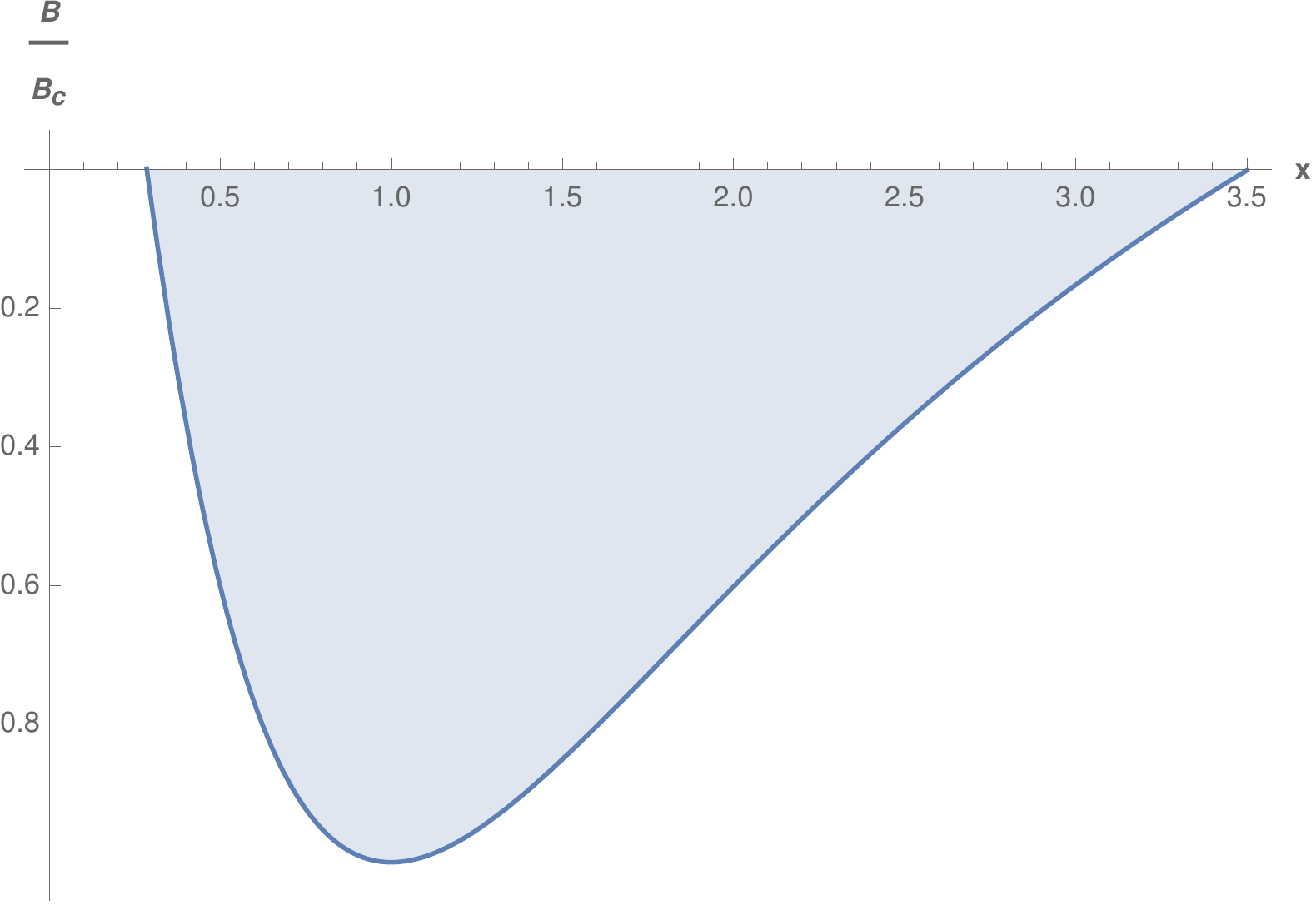}
\includegraphics[width=6cm]{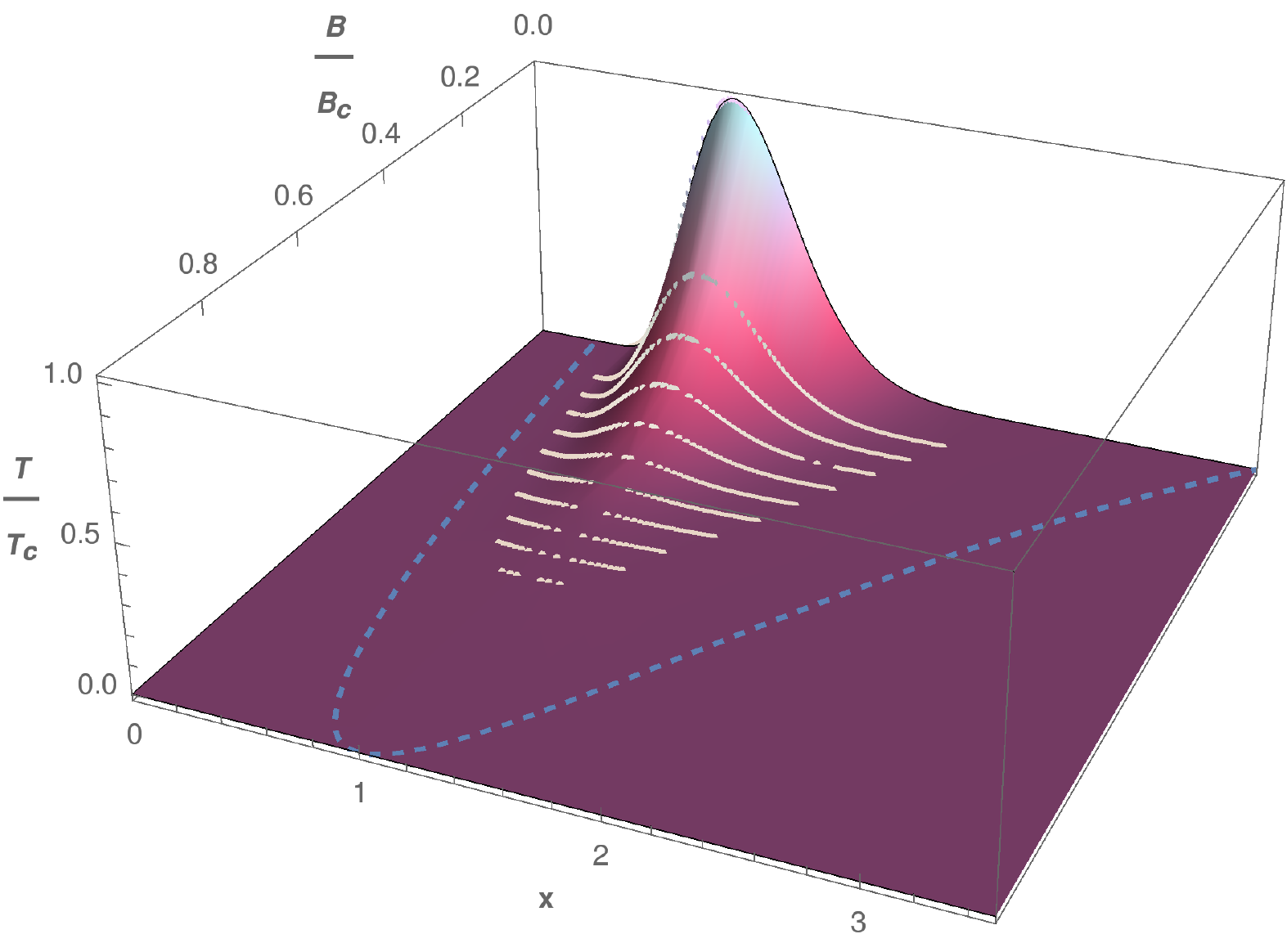}
\caption{Top: $(T,{\sf x})$ plane. Bottom: $(B,{\sf x})$ plane (left) and the complete $(B,T,{\sf x})$ diagram (right).
We have   $m^2=-0.4$, $q=\bar{q}=0.6$
and ${\sf y}=0$.  We are normalizing the plots with $T_c$, the maximal critical temperature for $B=0$, and $B_c$, the maximal magnetic field for $T=0$. In this case $B_c \simeq 7.764 \times 10^{10} T_c^2 $.}
\label{fig:PhaseDiagram_m2=-04}
\end{figure}

For the three particular examples discussed in the previous section, we have integrated numerically the equations of motions for generic values of the parameters $({\sf x},T,B)$ at fixed ${\sf y}$, obtaining the superconducting and the normal phases. The results are shown in Figs. \ref{fig:PhaseDiagram_m2=-2} to \ref{fig:PhaseDiagram_m2=-04}. As it can be seen from those figures, with the  model presented in section \ref{sec:HolographicTheory} one can realize diverse types of $({\sf x},T,B)$ phase diagrams by adopting different values for the parameters of the theory. In all the cases, the
numerical results are in agreement with the kind of behavior we have predicted analytically, in terms of the values of $q$ and $\bar q$.

In particular, as it is shown in the case $iii)$ of Fig.\ref{fig:Bxplanes}, it is possible to find values of the scalar charges $q$ and $\bar q$ and  squared mass $m^2$, such that a superconducting dome appears at intermediate values of the doping, which resembles real doped high $T_c$ superconductors.  The dome is damped as the magnetic field $B$ is increased, and disappears completely at a critical value of $B$ that depends on the doping and the parameter {\sf y}. As the temperature is increased, the dome of superconductivity retracts, but  at any constant $T$ plane the maximum of the critical magnetic field is always at the value ${\sf x}_{\sf max}=\bar q/q$, as derived analytically for $T=0$. The same is true  at any constant $B$ plane for the maximum on the critical temperature. This shows that for the studied masses, the value ${\sf x}_{\sf max}$ is the only stationary point on equations \eqref{1stLine}-\eqref{2ndline}. In other words, the particular ratio of chemical potentials ${\bar\mu}/{\mu} = {\bar q}/{q}$ favours the existence of the superconducting phase. It would be interesting to further understand this fact.

A phase diagram with the above mentioned generic features can be found for suitable values of $q$, $\bar q$ and $m^2$. Varying the value of ${\sf y}$ deforms the superconducting region for $B>0$, but do not change the range of doping in which the system is superconductor.

Following \cite{Kiritsis2016HolographicSuperconductivity}, we have adopted the quotient of the chemical potential of the two U(1) currents as a definition of a doping-like variable {\sf x}. In this holographic model the two U(1) charges are independently conserved, but a more realistic model could be achieved by introducing a mass term coupling the two U(1) fields. As another possible extension, we plan to investigate the magnetic field axis on the doped model in a generic region of couplings, allowing for non-minimal interaction of the superconducting condensate with the bulk gauge fields. In such case, the mapping to the undoped model does not seem immediate, and a different approach may be needed.
\section*{Acknowledgments}
This work was partially supported by Conicet grants PIP 0681, PIP 1109, and PUE~  ``B\'us\-que\-da de Nueva F\'\i sica'' and UNLP grants X850 and X791. The authors thank Marco Cerezo, Gast\'on Giordano, Ignacio Salazar-Landea, Carlos Lamas, Mauricio Matera and Guillermo Silva for discussions.  They also thank Daniele Musso and Dimitrios Giataganas for helpful comments.

\newpage

\bibliographystyle{JHEP}
\bibliography{references}
\end{document}